\documentclass[showpacs,superscriptaddress,amsmath,amssymb,nofootinbib,twocolumn,floatfix]{revtex4}

\usepackage{setspace}
\usepackage{graphicx}  
\usepackage{dcolumn}   
\usepackage{bm}        
\usepackage{rotating}
\usepackage{lipsum}
\usepackage{color}  
\usepackage{hyperref} 

\begin{document}


\title{Instrumented Baffle for the Advanced Virgo Input Mode Cleaner End Mirror}

\author{M. Andr\'es-Carcasona}
\author{O. Ballester}
\author{O. Blanch}
\affiliation{Institut de F\'\i sica  d'Altes Energies (IFAE), The Barcelona Institute of Science and Technology, Campus UAB, 08193 Bellaterra (Barcelona) Spain}
\author{J. Campos}
\affiliation{Instituto de \'Optica (IO-CSIC), 28006 Madrid, Spain} 
\author{G. Caneva}
\author{L. Cardiel}
\author{M. Cavalli-Sforza}
\affiliation{Institut de F\'\i sica  d'Altes Energies (IFAE), The Barcelona Institute of Science and Technology, Campus UAB, 08193 Bellaterra (Barcelona) Spain}
\author{P. Chiggiato}
\affiliation{CERN, 1211 Geneva, Switzerland}
\author{A. Chiummo}
\affiliation{European Gravitational Observatory (EGO), 56021 Cascina, Pisa, Italy}
\author{V. Dattilo}
\affiliation{European Gravitational Observatory (EGO), 56021 Cascina, Pisa, Italy}
\author{ J.A. Ferreira}
\affiliation{CERN, Geneva, Switzerland}  
\author{J.M. Illa }
\author{C. Karathanasis}
\author{M. Kolstein}
\affiliation{Institut de F\'\i sica  d'Altes Energies (IFAE), The Barcelona Institute of Science and Technology, Campus UAB, 08193 Bellaterra (Barcelona) Spain}
\author{M. Mart\'\i nez}
\affiliation{Institut de F\'\i sica  d'Altes Energies (IFAE), The Barcelona Institute of Science and Technology, Campus UAB, 08193 Bellaterra (Barcelona) Spain}
\affiliation{Instituci\'o Catalana de Recerca i Estudis Avançats (ICREA), 08010 Barcelona, Spain}
\author{A. Macquet}
\author{A. Men\'endez-V\'azquez}
\author{Ll.M. Mir}
\author{J. Mundet}
\affiliation{Institut de F\'\i sica  d'Altes Energies (IFAE), The Barcelona Institute of Science and Technology, Campus UAB, 08193 Bellaterra (Barcelona) Spain}
\author{A. Pasqualetti}
\affiliation{European Gravitational Observatory (EGO), 56021 Cascina, Pisa, Italy}
\author{O. Piccinni}
\author{C. Pio}
\author{A. Romero-Rodr\'\i guez}
\author{D. Serrano}
\affiliation{Institut de F\'\i sica  d'Altes Energies (IFAE), The Barcelona Institute of Science and Technology, Campus UAB, 08193 Bellaterra (Barcelona) Spain}

\date{\today}

\begin{abstract}

A novel instrumented baffle surrounding the suspended end mirror in the input mode cleaner 
cavity of the Virgo interferometer was installed in spring 2021. 
Since then,  
the device has been regularly operated in the experiment 
and the data collected show good stability.
The baffle will operate in the upcoming O4 observation run,
serving as a demonstrator of the technology designed to instrument the baffles in front of the main mirrors
in time for O5.
In this paper we present a detailed description of the baffle design,
including mechanics,
front-end electronics,
and data acquisition
as well as optical and vacuum compatibility tests,
calibration and installation procedures,
and performance results.

\end{abstract}

\maketitle

\section{Introduction}
\label{sec:intro}

One of the main challenges of current interferometers is the suppression of stray light (SL)
to minimise noise due to possible re-coupling with the main beam to typically one order of magnitude below the target sensitivity of the interferometer. In particular, the SL inside the interferometer arms mainly originates from 
light scattered off the mirror surfaces,  and also includes diffractive contributions and edge effects due to limiting apertures and the clipping of the main fundamental mode inside the cavity.   
The approach taken by Advanced Virgo~\cite{Acernese_2014} in the first three observation runs was to equip the interferometer with low-reflective, 
low-scattering baffles,
either connected to ground in the vacuum tubes or suspended around the mirrors, 
to absorb this light.
Although these baffles served their purpose to mitigate SL below the noise floor,
there was no real control or monitoring of the SL distribution.
In addition, 
Virgo interferometer arms do not have active sensors in the vicinity of the test masses to assist in the pre-alignment of the interferometer. Altogether, this motivated the need for upgrading the experiment by installing instrumented baffles surrounding the test masses in time for the O5 observation run~\cite{Virgo:2019juy}. 

\begin{figure}[htb]
\begin{center}

 \includegraphics[width=0.45\textwidth]{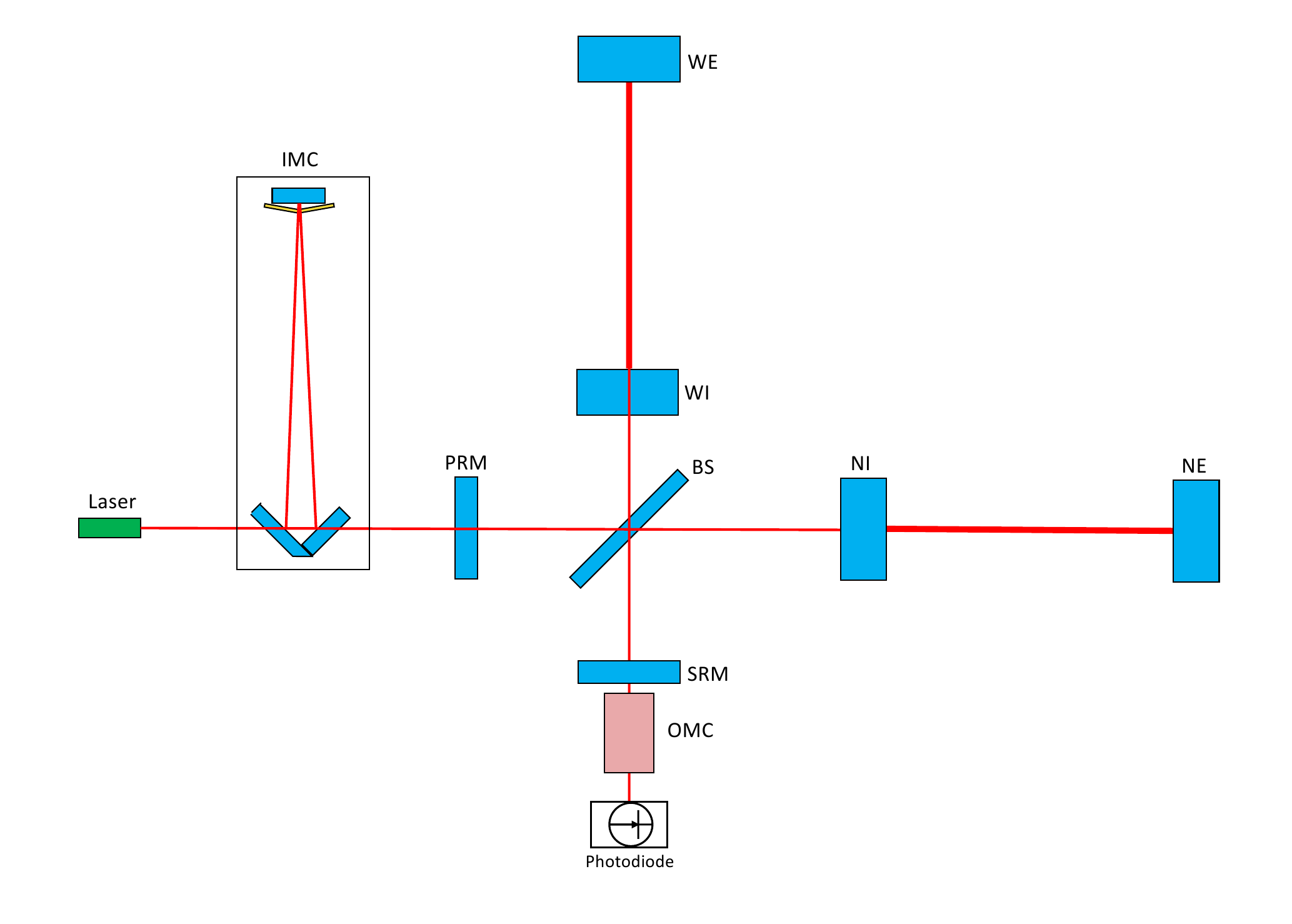}

\end{center}
\caption{\small
Simplified,
not to scale, 
optical layout of the Advanced Virgo interferometer. 
Each interferometer arm is formed by an input mirror (North input (NI) or West input (WI)) 
and an end mirror (North end (NE) or West end (WE)).
The IMC cavity,
located in front of the interferometer arms,
is shown highlighted.
}
\label{fig:VirgoSketch}
\end{figure}

In April 2021, 
a first instrumented baffle was installed surrounding the suspended mirror of the input mode cleaner (IMC) cavity at Virgo, 
to serve as a demonstrator of the technology designed to instrument the baffles in the interferometer arms in the near future.
The IMC is a triangular cavity formed by two flat mirrors that form a dihedron and a suspended curved mirror,
whose purpose is to stabilize the frequency of the laser~\cite{Acernese_2014}. 
A sketch of the interferometer is shown in Fig.~\ref{fig:VirgoSketch}.

The instrumented baffle has been operated since its installation with success,  
demonstrating its functionality.  
The measured light distribution is in good agreement with SL simulations, 
dominated by the mirror surface maps inside the cavity~\cite{Ballester_2022}.  
The IMC instrumented baffle will be operating in the upcoming O4 observation run, 
and we expect that the information provided by the baffles will not only improve 
the understanding of the SL distribution at low angles in the interferometer, 
but also detect the appearance of higher order modes,
which appear as modified patterns in the SL distribution,
open the possibility of monitoring the contamination of the mirror surfaces that leads to low-angle scattering,
and facilitate a more efficient pre-alignment and fine-tuning of the parameters of the interferometer after shutdowns and during operations.
 
The instrumented baffle design requirements,
the different steps of the assembly,
the calibration of the sensors in the laboratory,
and the installation in the IMC end mirror cavity are described in the following sections,
together with the first results on the baffle operation.

\section{Baffle Requirements}
\label{sec:design}

The starting point of the design was the replacement of the
existing baffle by one equipped with photo-sensors,
keeping its weight and mass distribution as close as possible to the original ones 
to avoid having to re tune the payload after the baffle installation.
The material of the existing baffle was AISI 304L stainless steel,
with reflectance and total integrated scattering measured values typically around 0.5\% and 300 ppm,
respectively.
The new baffle had to meet the same optical requirements,
and all its elements had to be certified for ultra high vacuum (UHV) 
conditions.~\footnote{
The IMC operates in a high vacuum (HV) regime.
However, 
the requirements for this baffle were set to comply with the UHV environment in which the baffles 
surrounding the test masses in O5 will be operated.}

\begin{figure}[htb]
\begin{center}
 \includegraphics[width=0.3\textwidth]{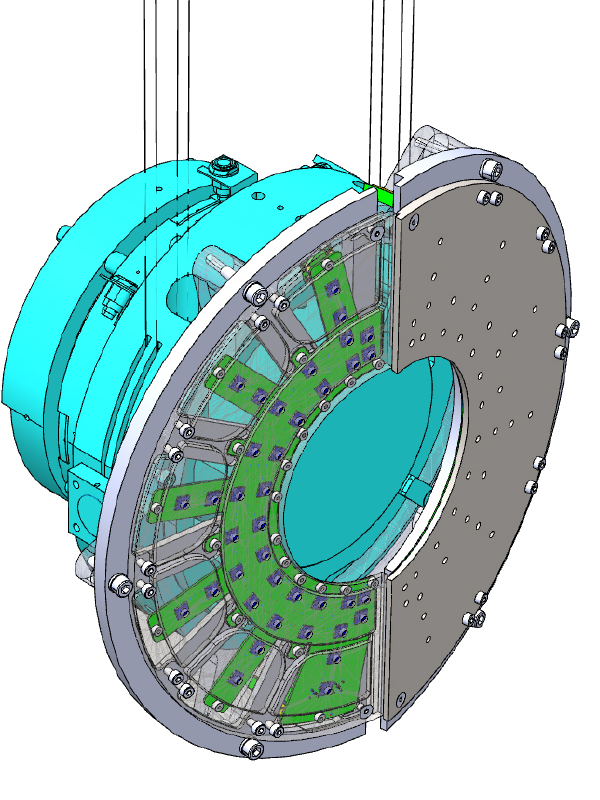}
\end{center}
\caption{\small
Design of the instrumented baffle for the end mirror of the IMC. 
The laser light reaches the photo-sensors through holes in the stainless steel plates.
On the right side, 
the fully assembled baffle is shown, 
while on the left side, 
the top cover has been removed to reveal the PCB board equipped with photo-sensors.
}
\label{fig:baffleDesign}
\end{figure}

Since most of the light is expected to be near the inner edge of the baffle~\cite{Romero:2020ovg},
the majority of the sensors are placed in that area,
on two large gold-plated polyimide-based printed circuit boards (PCBs).
The sensor signals are processed by 16 ADCs,
which send their data to the data acquisition (DAQ) server through either a serial or a wireless system.
Because of the UHV environment, 
heat dissipation will occur mainly through radiation. 
To favor this process, 
the PCBs are thermally connected to the metallic structure of the baffle, 
thus avoiding the concentration of heat in critical semiconductor components.
Figure~\ref{fig:baffleDesign} shows the design of the instrumented baffle for the end mirror of the IMC.

\section{Baffle Mechanics}
\label{sec:mechanics}

\begin{figure}[htb]
\begin{center}
 \includegraphics[width=0.4\textwidth]{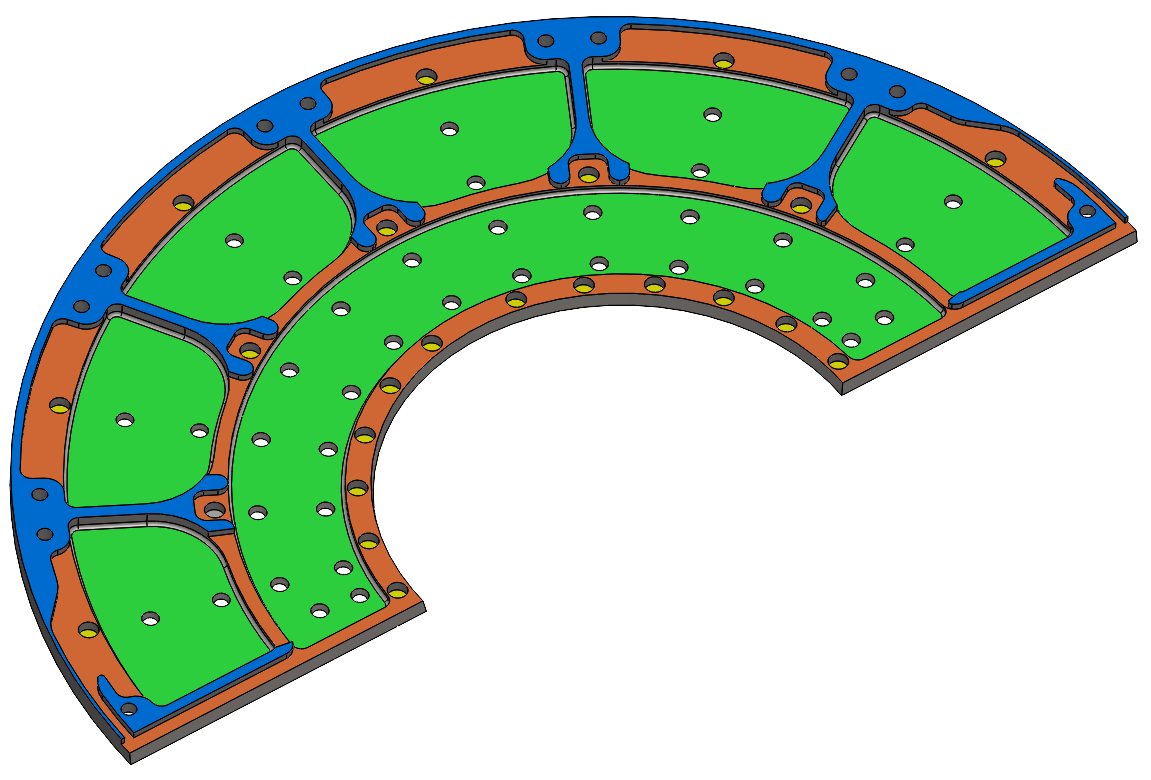}
\end{center}
\caption{\small
Rear side of the baffle.
The thickness of the green, orange, and blue areas are 2, 4, and 6 mm, 
respectively.
}
\label{fig:baffleThickness}
\end{figure}

The baffle has an inner (outer) radius of 70 (175)~mm and is divided into two halves, 
each tilted nine degrees with respect to the plane normal to the nominal direction of the laser beam to prevent full back reflections into the IMC cavity. 
The rear of the baffle is pocketed to accommodate the electronics board while meeting weight and center of mass requirements and maintaining necessary stiffness. The thickness of the baffle is 2~mm except in the rings and spokes, 
where thicknesses of 4 and 6~mm alternate, 
as shown in Fig.~\ref{fig:baffleThickness}.

\begin{figure}[htb]
\begin{center}

 \includegraphics[width=0.4\textwidth]{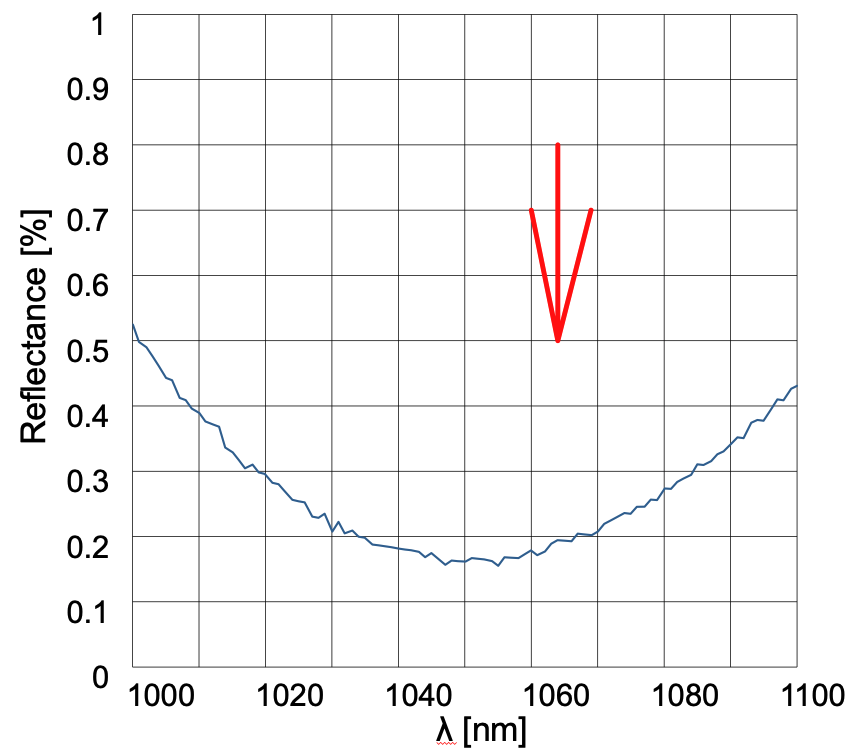}

\end{center}
\caption{\small
Report of the metallic coating provided by Optimask~\cite{optimask}.
The arrow shows the reflectance for 1064~nm incident light at an angle of eight degrees.
}
\label{fig:coating}
\end{figure}

The baffle material is AISI 304L stainless steel with super mirror finish number 8,
obtained by lapping together with a chemical-mechanical polishing technique,
fulfilling the required average roughness with a value of Ra~=~0.0069~$\mu$m~\cite{sensofar} 
and a total integrated scattering between 500 and 800 ppm.
The outer border of the baffle was machined by milling,
whereas the required sharpness of the inner border was obtained with wire-electrical discharge machining (EDM),
reaching a radius of curvature of approximately 10~$\mu$m. 
This inner border was cut at an angle of ten degrees to further avoid back reflections.
The relative magnetic permeability of the baffle, $\mu_r$, was measured to be 1.08.
The surface of the baffle facing the cavity was covered with a CrCrO2 anti-reflective coating 
that met the requirement on the reflectance to be smaller than 0.5\% for incident light of 1064~nm at an angle of eight degrees,
as can be seen in Fig.~\ref{fig:coating}.

\begin{figure}[htb]
\begin{center}

 \includegraphics[width=0.4\textwidth]{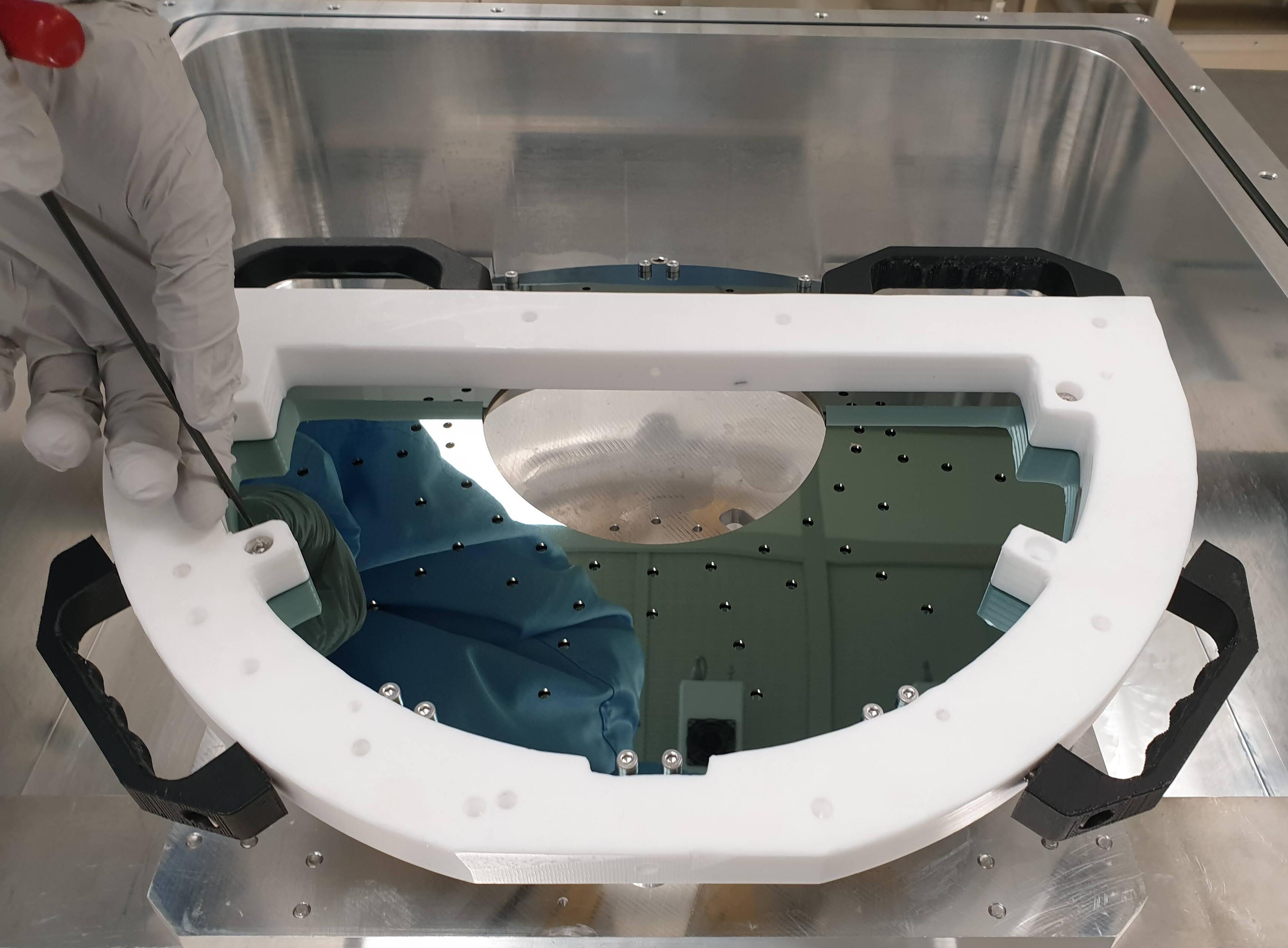}

\end{center}
\caption{\small
Half-baffle being bolted to the PTFE support with handles.
}
\label{fig:baffleMold}
\end{figure}

To protect the polished surface after the coating process until the final installation 
of the baffle in the IMC end mirror cavity,
each half-baffle was fixed, 
through a subset of the holes intended to bolt the baffle to the payload, 
to a polytetrafluoroethylene (PTFE) support with handles to prevent manipulating it directly with the hands
during the different stages of assembly and installation,
as can be seen in Fig.~\ref{fig:baffleMold}.

\begin{figure}[htb]
\begin{center}

 \includegraphics[width=0.4\textwidth]{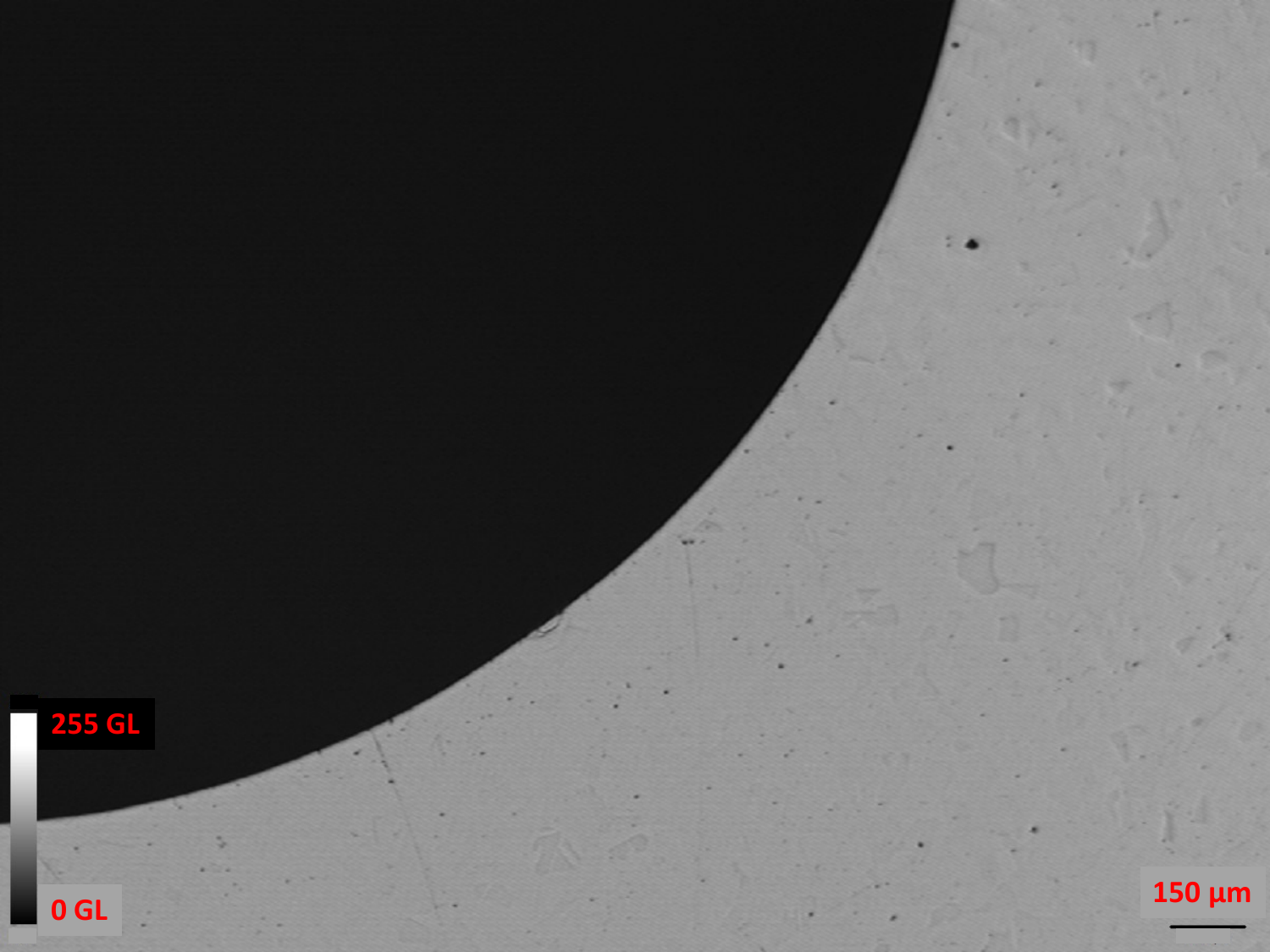}

\end{center}
\caption{\small
Edge of a hole seen with a $5\times$ magnification lens.
}
\label{fig:holeEdge}
\end{figure}

The light reaches the photo-sensors at the back of the baffle through conical holes
with an opening angle of twelve degrees,
such that the light does not resolve the sensor edges and the scattering of light within the hole geometry is minimised.
The holes are centered at the sensors and have a diameter of 4~mm at the front of the baffle.
They were machined by helical interpolation milling to reach the required sharpness precision,
visible in Fig.~\ref{fig:holeEdge}.

Table~\ref{tab:baffleWeight} shows the weights of the different components of the instrumented baffle,
which results in a baffle 160~g lighter than the previous one.
The new center of mass moved away (-0.34, 0.82, 2.02)~mm with respect to the non-instrumented baffle. 
These small differences in weight and center of mass with respect to the old baffle were easily accommodated by tuning the new payload.

\begin{table}[htb]
\caption{\small
Weight of different components of the instrumented baffle compared to the previous non-instrumented baffle.
}
\begin{center}
\begin{small}
\begin{tabular}{l r l r} 

\hline
\multicolumn{2}{c}{O3 (plain) baffle} & \multicolumn{2}{c}{O4 (instrumented) baffle} \\ 
\hline
Components       & Weight (g)  & Components         & Weight(g)           \\
\hline
Stainless steel  & 1887.47     & Stainless steel    & 753.1 + 751.6       \\
                 &             & Screws             & 0.17 $\times$ 46    \\
                 &             & PCBs               & 95.77 + 98.03       \\
                 &             & Antennae           & 11.14               \\
                 &             & Connectors         & 5.00 $\times$ 2     \\
                 &             & Cables             & 0.56                \\
\hline                 
Total            & 1887.47     & Total              & 1728.02             \\
\hline
\multicolumn{4}{c}{Difference = 159.45 g}  \\
\hline

\end{tabular}
\end{small}
\end{center}
\label{tab:baffleWeight}
\end{table}

\section{Infrared Sensors}
\label{sec:sensors}

According to simulation, 
$2\times10^{-4}$ ($1\times 10^{-6}$)~W/cm2 per watt of laser input power would reach a sensor located on the inner (outer) edge of the baffle, 
and a granularity of approximately 100~mm$^2$ would be sufficient to allow detection of changes in the light pattern caused by small misalignments of the IMC cavity~\cite{Romero:2020ovg}.
Thus, 
sensors of this size that could receive of the order of a few~mW of power, 
with a temporal precision of around 1~Hz and optical properties similar to those of the baffle,
i.e.,
a very low reflectance to minimize the light not hitting the main mirror to be reflected back,
as well as low integrated scattering are needed.
Finally, 
because of the UHV environment the sensors would be working in,
the amount of outgassing has to be below $10^{-4}$~mbar l/s.
A specific research and development project with the Hamamatsu Photonics company~\cite{hamamatsuCompany} 
to develop and produce this type of sensors was established 
and the proposed solution is based on the S13955-01 photodiode, 
with the under filling removed~\cite{hamamatsu}.
Table~\ref{tab:sensors} summarises the properties of the final version of the sensors.

\begin{table}[htb]
\caption{\small
Properties of the silicon photodiodes developed by the Hamamatsu Photonics company for the baffle instrumentation project.
}
\begin{center}
\begin{small}
\begin{tabular}{l r} 

\hline
\multicolumn{2}{c}{Sensor characteristics}   \\
\hline
Dimensions            &    $7.37 \times 7.37$~mm$^2$  \\
Sensitive area        &    $6.97 \times 6.97$~mm$^2$  \\
Operation temperature & $-40^{\circ}$C to $100^{\circ}$C     \\
Reflectance          &    1.2 to 1.8\% at $4^{\circ}$ incident angle  \\
Photo-sensitivity     &    660~mA/W               \\

\hline

\end{tabular}
\end{small}
\end{center}
\label{tab:sensors}
\end{table}

\begin{figure}[htb]
\begin{center}
    
 \includegraphics[width=0.4\textwidth]{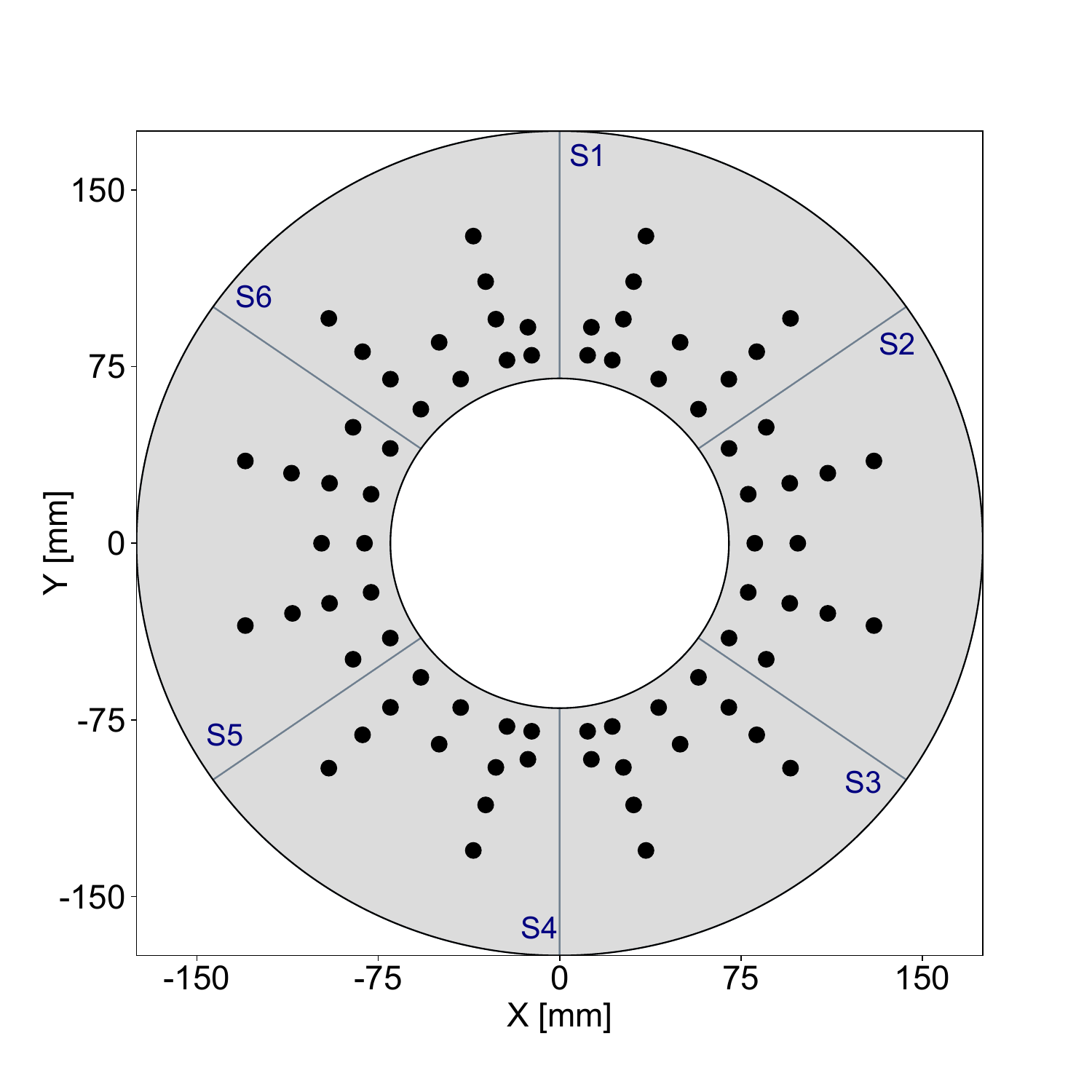}

\end{center}
\caption{\small
Sketch of the location of the photo-sensors in the baffle in the XY plane.
}
\label{fig:sensorsDistribution}
\end{figure}

Based on the simulation predictions,
the layout of the sensors consists of two concentric rings at radii of 81 and 98 mm, respectively, 
with additional sensors distributed across the baffle at larger radii and equally-spaced azimuthal angles,
for a total of 76 sensors,
as shown in Fig.~\ref{fig:sensorsDistribution}.

\section{Optical Tests}
\label{sec:optests}

Dedicated measurements of the reflectance and total induced scattering of the photo-sensors took place at 
Instituto de \'Optica - Consejo Superior de Investigaciones Cient\'{\i}ficas (IO-CSIC) and 
European Gravitational Observatory (EGO). 
The different versions of sensors provided by the Hamamatsu Photonics company presented an improved performance in terms of reduced reflectance at 1064~nm,
which decreased from about $50\%$ in the first version to about 1.2\% to 1.8\% in the sensors finally installed in the detector. 
Further attempts to reduce the total reflectance of the sensors translated into lower reflectance numbers at the cost of loss of linearity under the reduced bias voltage applied. 
Since maintaining linearity in the detector response is a priority we 
finally adopted the so-called version 3 sensors with a reflectance slightly above the 1\% level.  

The total integrated scattering (TIS) was measured in realistic conditions at EGO, 
taking into account the presence of the baffle and the conical shaped holes in front of the sensors.  
Since part of the light is retained by the mechanical structure, 
the scattered light off the sensor is confined to a cone of twelve degrees centered at the sensor 
and the corresponding TIS will be reduced. 

\begin{figure}[htb]
\begin{center}

 \includegraphics[width=0.2\textwidth]{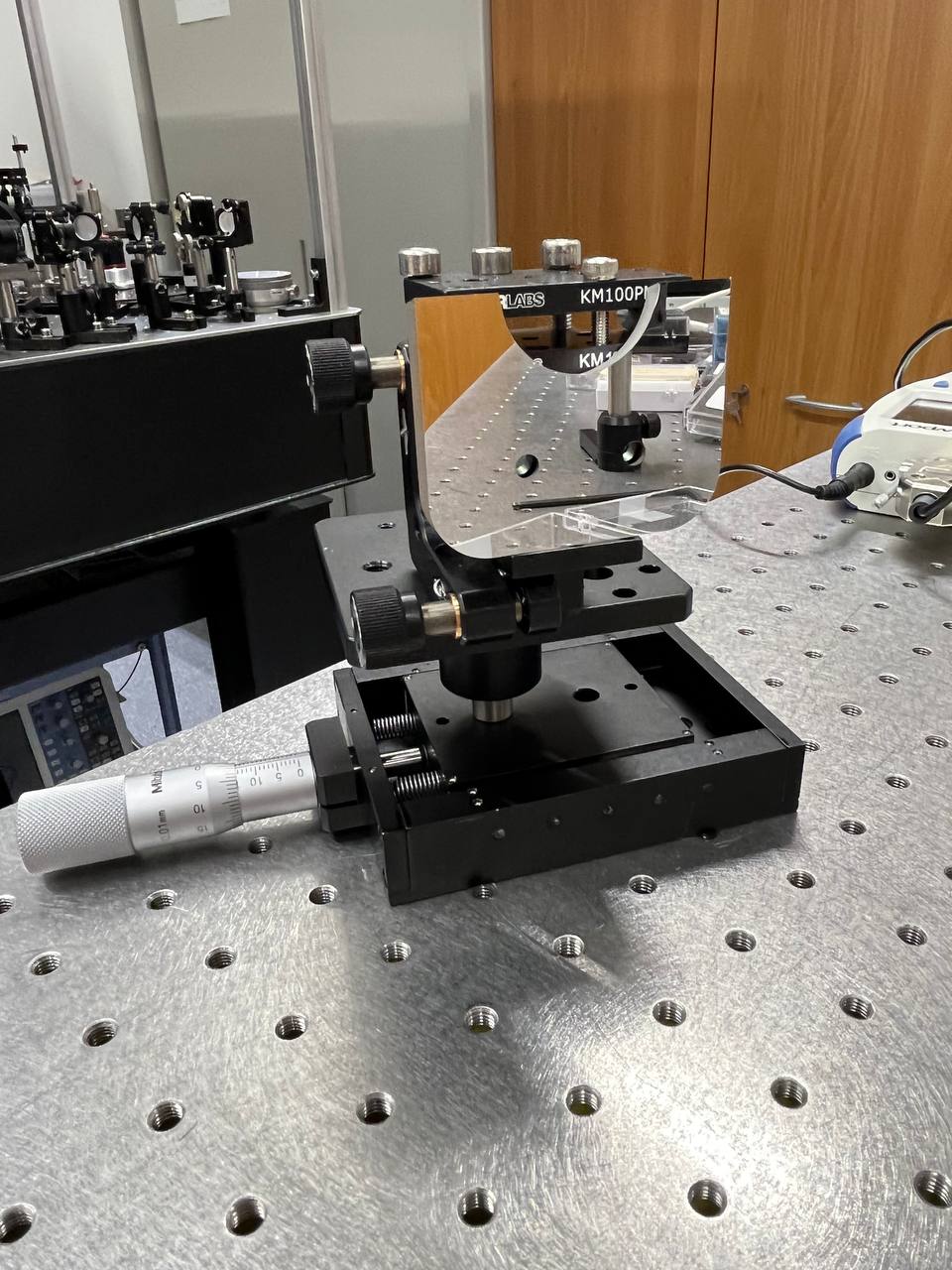} 
 \includegraphics[width=0.2\textwidth]{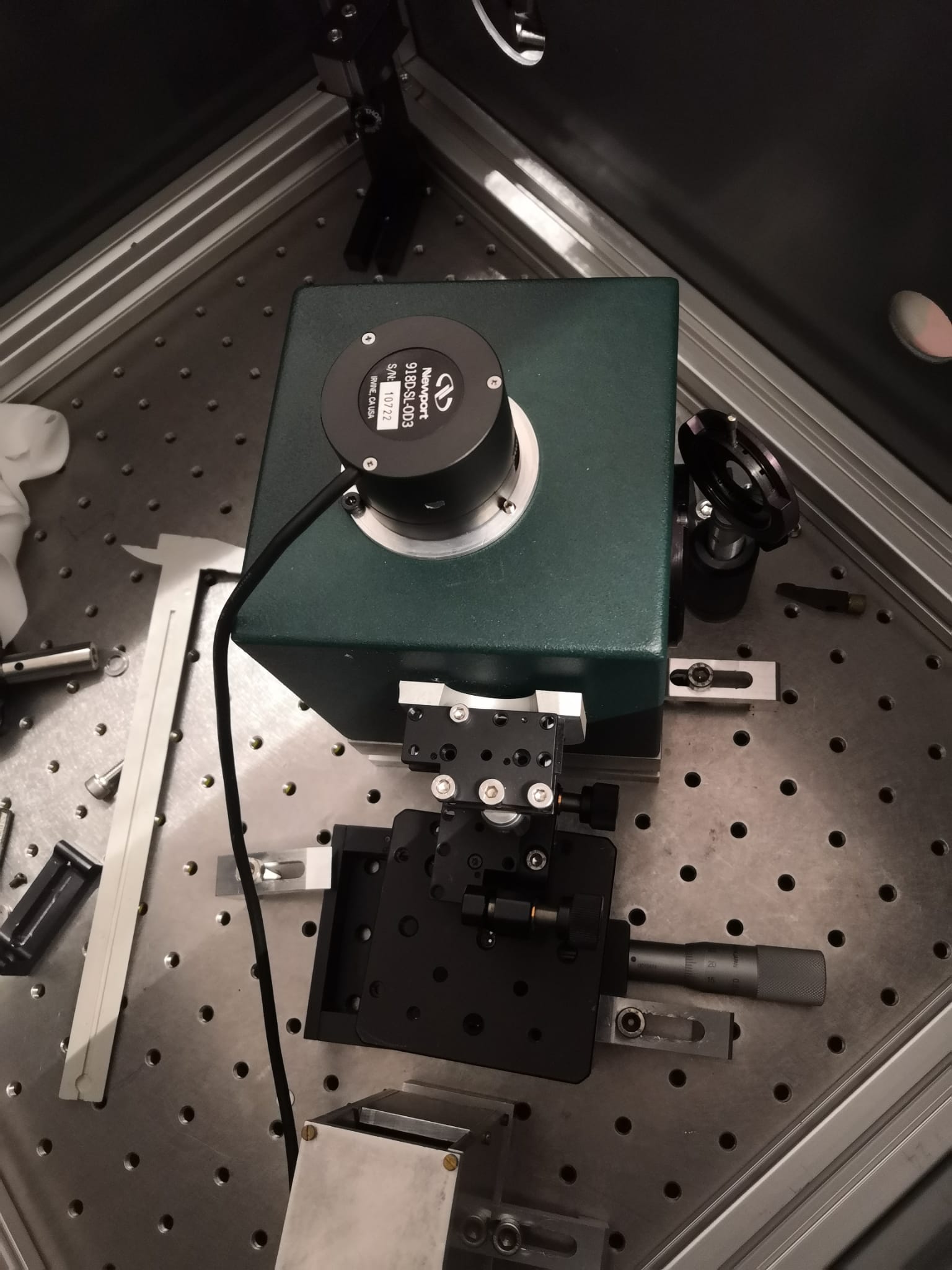}  

\end{center}
\caption{\small
Optical setup at EGO to measure the sensor's TIS.
(left) In vertical position a small sample with two holes and photodiodes behind them, 
with the same geometry as those of the instrumented baffle, attached to a support.
(right) Sample placed in front of the integrating sphere.
}
\label{fig:opticalTestSetup}
\end{figure}

The optical setup consisted of a collimated 1064~nm laser beam, 
which was expanded and linearly polarised through a series of lenses, 
wave plates, and polarising beam splitters, 
and finally directed toward an integrating scattering sphere with five ports. 
The test sample was mounted near the port orthogonal to the beam, 
and the sample was tilted slightly to direct the specularly reflected light off the sample out of the integrating sphere and into a beam dump. 
This ensured that the light measured by means of a photodiode mounted on the port located at the top of the box shielding the integrating sphere 
and a power meter was due to scattering only. 
An ideal diffuser with a Lambertian scatter profile was used as reference,
and the ratio between the power scattered off the reference surface, 
which has a scattering efficiency of 100 \%, 
and the power scattered off the sample under investigation with the same incident power was measured. 
A measurement with the probe port empty and the beam dumped into a very efficient beam dump located about 20~cm away provided a measurement of the noise floor due to spurious stray light from other sources. 
This noise floor, 
typically less than 1 ppm, 
was a systematic bias subtracted from the final measurement. Figure~\ref{fig:opticalTestSetup} shows the test sample and the integrating sphere.

The level of scattering at the bulk of the mechanics is below 600~ppm 
and indicates the high quality of the metal polishing. 
The scattering increases significantly at the edge of the baffle hole,
reaching values of 30,000~ppm or 3\% and peaks when the laser light illuminates the photo-sensor.
This is attributed to a number of factors related to the low quantum efficiency of silicon to 1064~nm light. 
That light is then scattered by the existing metal layer at the back side of the silicon chip and/or at the ceramic packaging layer, 
which are optically untreated surfaces.
Although these values are a priori high, 
the induced effect on the operations of the IMC cavity is very small, 
since the scattered light off the IMC baffle is redirected away from the main stream thanks to the nine-degrees inclination of the baffle with respect to the direction normal to that of the incident beam in the cavity.

\section{Front-End Electronics}
\label{sec:fe}

Various external constraints,
including the UHV environment and 
the number and length of the available readout cables, 
as well as the sensors dynamic range, 
are considered in the design of the baffle front-end electronics.
Table~\ref{tab:electronicRequirements} summarises the requirements that constrain this design.

\begin{table}[htb]
\caption{Summary of the electronic requirements of the front-end electronics.}
\begin{center}
\begin{small}
\begin{tabular}{l  r} 

\hline
\multicolumn{1}{c}{Parameter}  & \multicolumn{1}{c}{Value}  \\ 
\hline
Communication system     &  Serial and wireless \\
Heat dissipation         &  2.5~W maximum by radiation \\
Dynamic range            &  20~mW for a 1064~nm wavelength \\
Resolution               &  0.125~mW  \\
Reading rate             &  2~Hz \\
Power / readout cables   &  Eight 15 m long \\
Cables type              &  \shortstack{Shielded twisted pairs \\ American wire gauge 24}  \\
Cables permittivity      &  5.6 nF/m  \\
\hline
\end{tabular}
\end{small}
\end{center}
\label{tab:electronicRequirements}
\end{table}

The baffle is equipped with two PCBs,
both with the capability to host the control and communication system,
which is implemented only in one of them.
The power connector is also located on this control PCB,
and the interconnection between the two PCBs is made through two 
RSP-175312-08 coaxial cables,
which are especially produced at the factory without any tag to avoid UHV incompatibilities, 
with a length of 145~mm and PTFE and gold compounds, 
suitable for UHV.

All electronic components can be powered at 3.3~V, 
which coincides with the bias voltage of the photodiodes, 
thus avoiding the use of high consumption regulators,
which would heat up the electronics,
as well as DC/DC converters, 
which would most likely generate outgassing.
To separate the polarisation of the photodiodes from the power supply of digital components,
a filter is implemented in each of the devices.
A voltage surge protection is implemented with a Zener diode with Vz~=~4.2~V, 
and the current protection is limited to 400~mA in the power supply,
which corresponds to a maximum system power of 1.3~W.
The communication and bias connectors are LEMO EZG.00.304.CLN, 
four-wire coaxial with polyether ether ketone (PEEK) insulation, 
ideal for UHV. 

\begin{figure}[htb]
\begin{center}

   \includegraphics[width=0.4\textwidth]{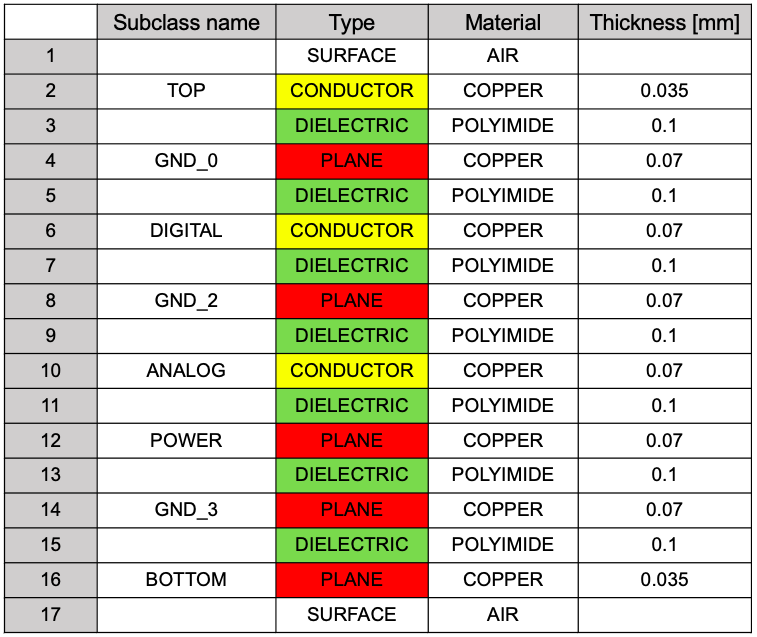} 

\end{center}
\caption{\small
Cross-section of the PCB showing the type,
material and thickness of the different layers.
}
\label{fig:pcbXsection}
\end{figure}

Polyimide is chosen as the insulating material for the PCBs, 
due to its low out-gassing and ease in expelling moisture compared to FR4. 
This choice limits the construction temperature to $200^{\, \circ}$C, 
which complicates the assembly of the components, 
but the benefits of being able to comply with a UHV environment justify its use. 
The thickness of the inner copper layers is fixed to be 70~$\mu$m. 
This massive design should be able to stop the few photons that could leak through the openings in front of the photodiodes. 
The area of the copper layers directly below the photodiodes does not have any etching to preserve the metal casing.
The outer layers are kept at 35~$\mu$m to better control the impedance of the strip lines, 
which is essential in lines related to communication, 
both serial and wireless.
To increase the heat dissipation, 
a gold-nickel alloy coating with much better thermal conductivity than copper was applied to the outer layers.
An eight-layer stack is adopted, 
which in addition to forming an even greater barrier for photons, 
allows an optimal separation between analog and digital signals, 
completely isolating them and thus minimizing the electronic noise that could be induced.
Figure~\ref{fig:pcbXsection} shows the cross-section of the PCB.

\begin{figure}[htb]
\begin{center}

   \includegraphics[width=0.4\textwidth]{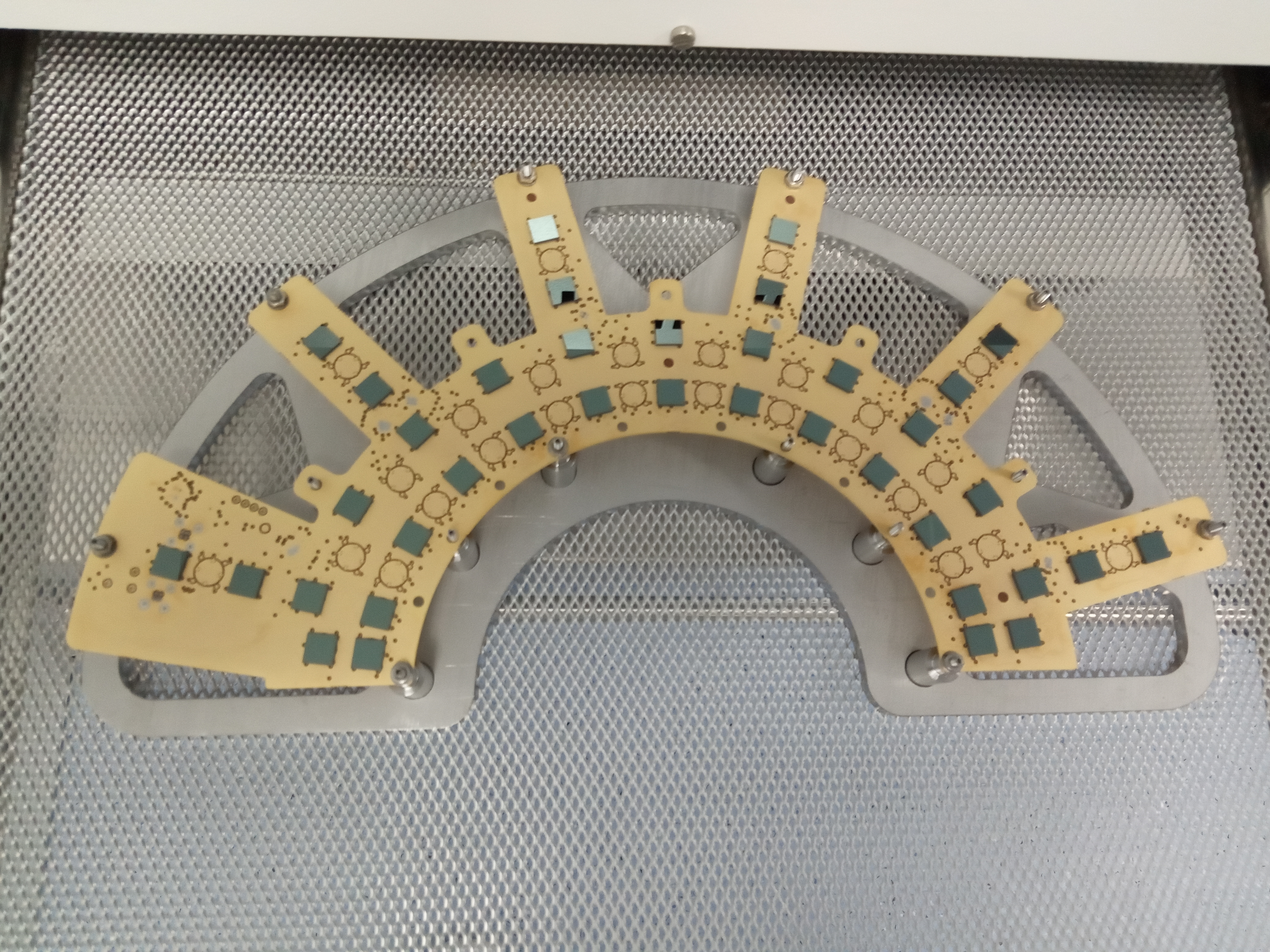}  \\
   \includegraphics[width=0.4\textwidth]{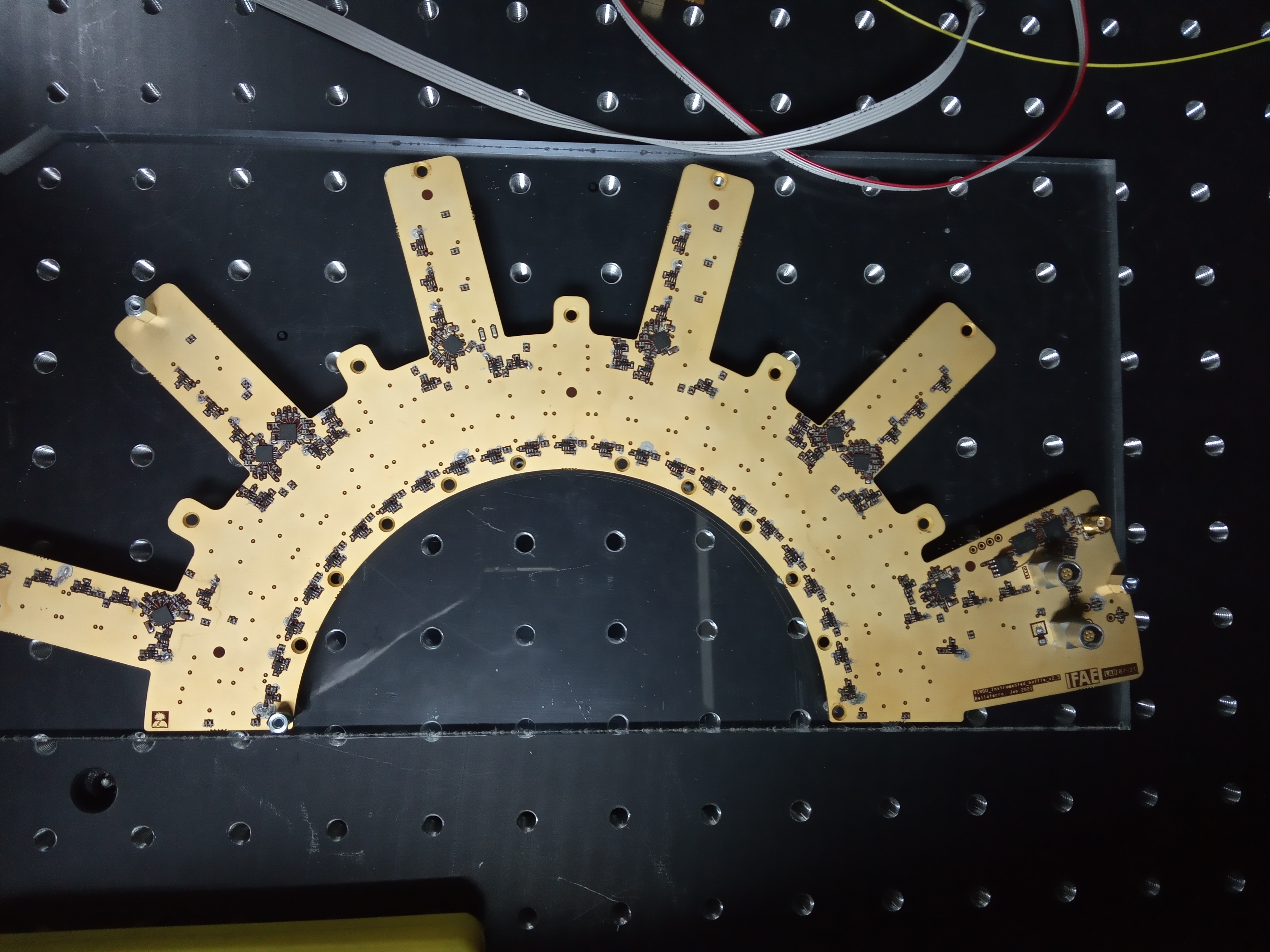} 

\end{center}
\caption{\small
(top) Top layer of the PCB containing the photodiodes.
(bottom) Bottom layer of the PCB, which contains all components except the photodiodes.
}
\label{fig:pcbDesign}
\end{figure}

\begin{figure}[htb]
\begin{center}

   \includegraphics[width=0.4\textwidth]{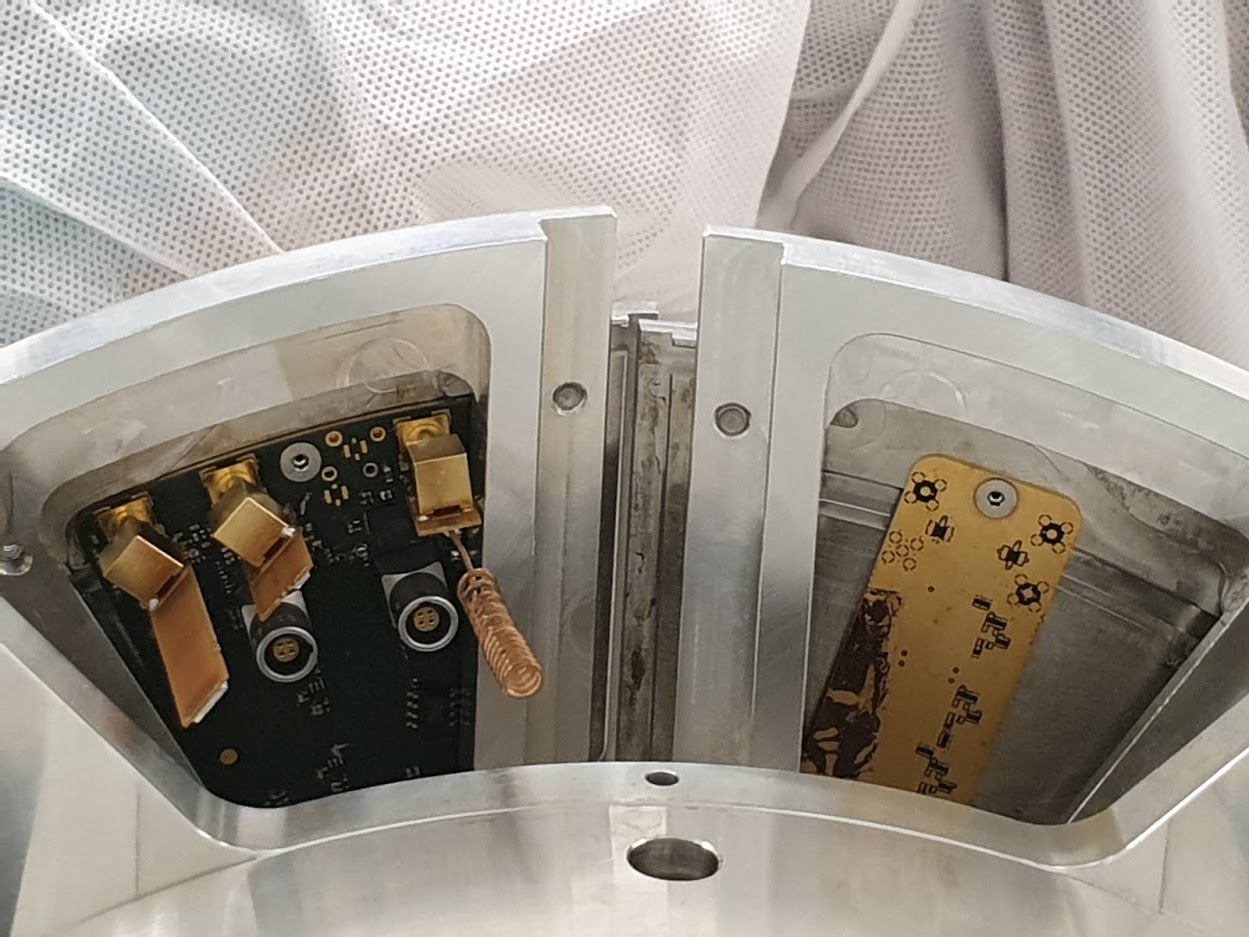} 

\end{center}
\caption{\small
Antennae for the wireless communication connected perpendicularly to the surface of the PCB.
}
\label{fig:antennae}
\end{figure}

The shape of the PCBs adapts to the supporting structure keeping its size to the minimum,
with all the electronic components placed in the free spaces of one side of the PCB, 
and the sensors placed on the other,
as shown in figure~\ref{fig:pcbDesign}.
Figure~\ref{fig:antennae} shows how the antennae for the wireless communication are not positioned on the PCB,
but attached perpendicularly to the surface of the board with a connector
to prevent the metallic structure of the baffle from having a Faraday box effect
that could interfere with communication.

\begin{figure}[htb]
\begin{center}

   \includegraphics[width=0.2\textwidth]{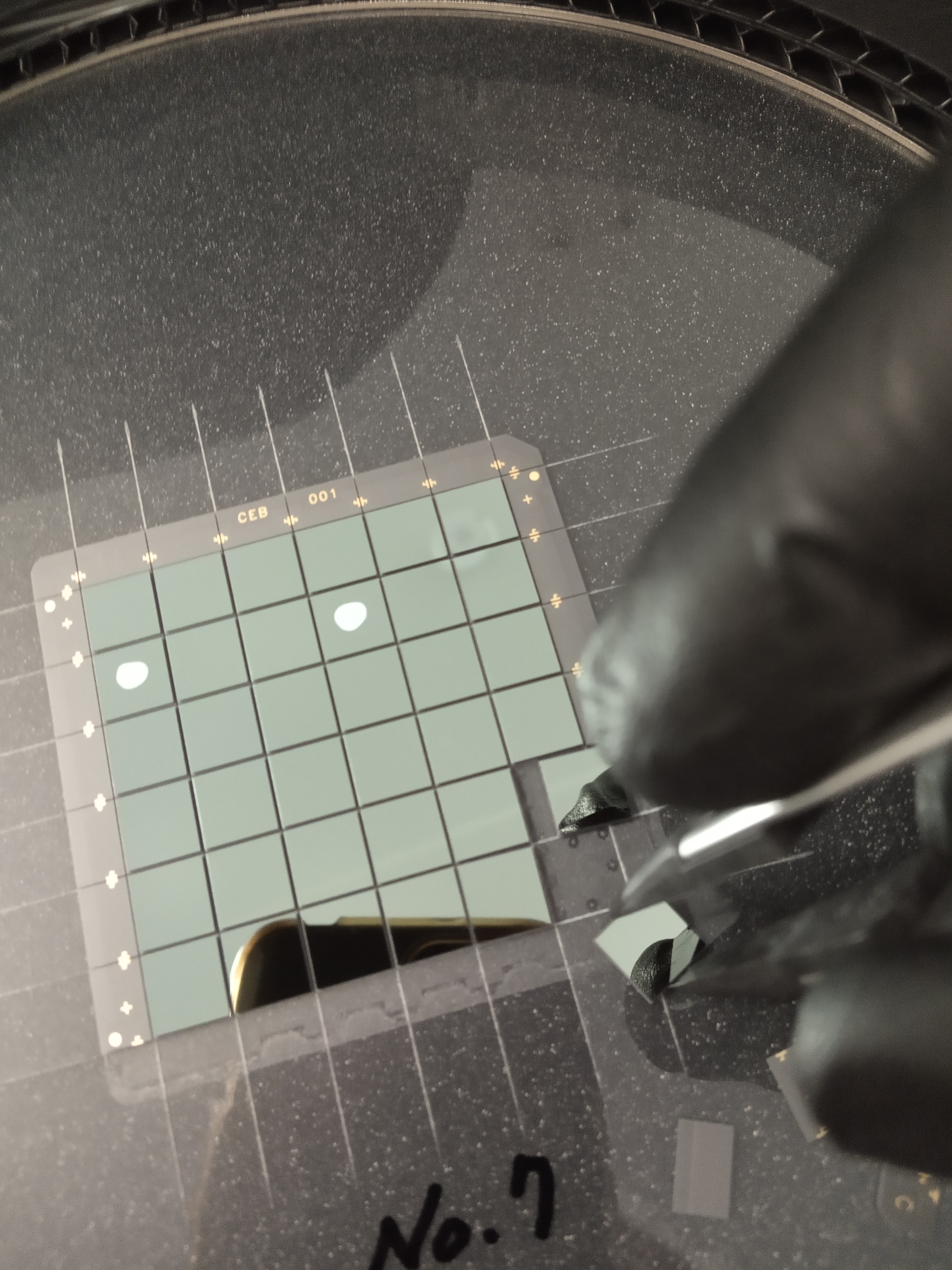} 
   \includegraphics[width=0.2\textwidth]{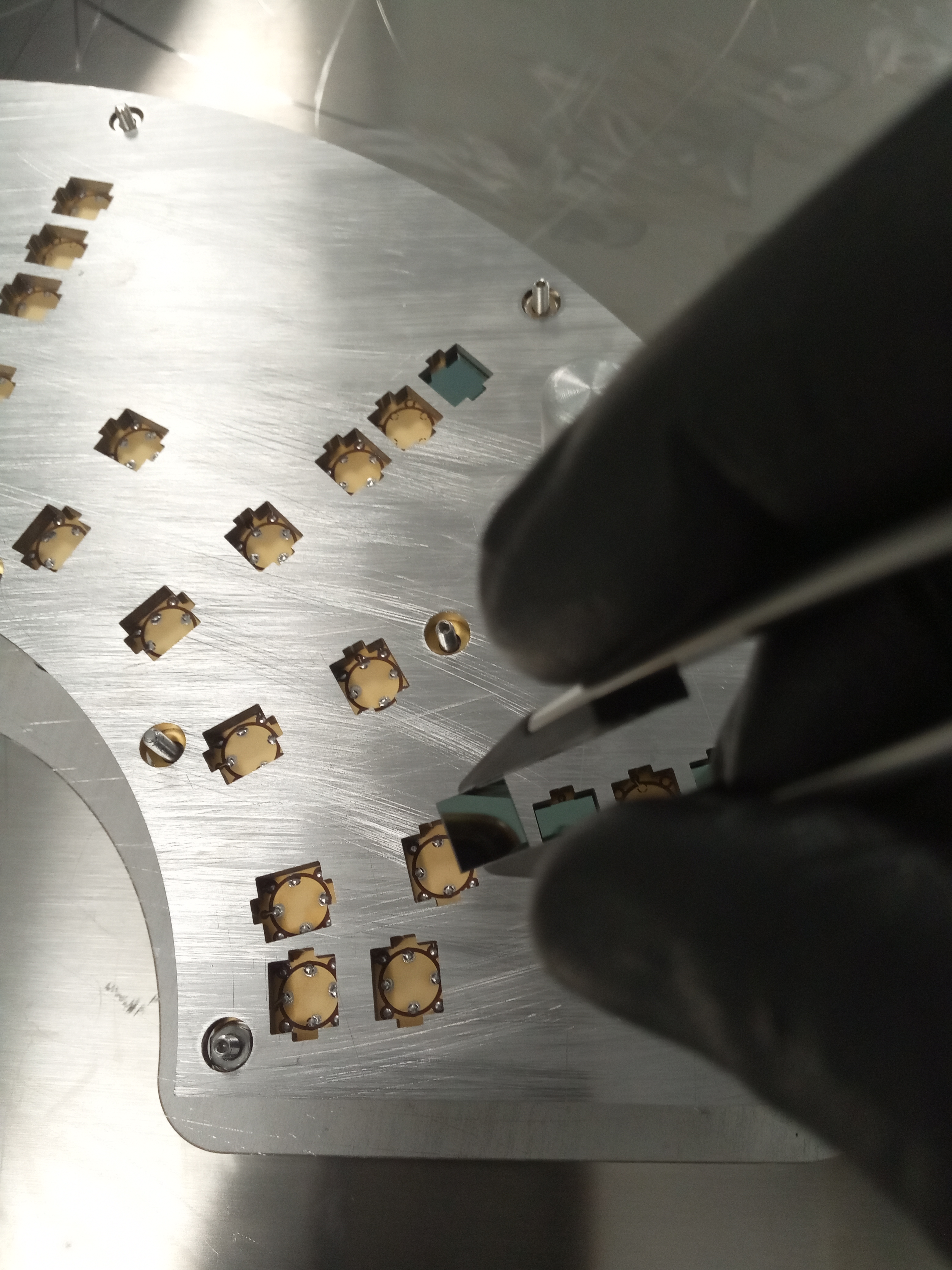} 

\end{center}
\caption{\small
(left) Photo-sensor being removed from the wafer and 
(right) placed on the PCB with the help of a template.
}
\label{fig:sensorTemplate}
\end{figure}

The special characteristics of these PCBs, 
without resin and with a gold cover,
together with its thick copper/polyimide structure,
made soldering in a FT05.B reflow oven of size 350~mm~$\times$~400~mm impossible.
Thus, 
the assembly proceeded in two steps:
First,
the bottom layer that contains all the components except the photodiodes was assembled in a vapor phase oven,
using ALPHA OM-353 SAC305 88.5-4-M20 solder paste, 
a 96.5Sn3.0Ag0.5Cu alloy completely lead free.
In a second step and after cleaning the boards, 
the photodiodes were soldered by hand with the help of
carbon-tipped tweezers and
a custom template on the top layer of the boards,
as shown in Fig.~\ref{fig:sensorTemplate}. 
Indalloy 1E solder paste was used in this case, 
with a eutectic composition of 50In50Ag alloy and a melting temperature 
of $118^{\, \circ}$C.
A reflow oven was used with a maximum temperature curve of $250^{\, \circ}$C, 
which guaranteed a uniformity of $160^{\, \circ}$C throughout the PCB, 
sufficient for the correct soldering of the photodiodes without affecting the other face. 
This second operation took place in the clean room to avoid contamination of the anti-reflective surface of the sensors.

The PCBs were cleaned twice,
first,
as mentioned above, 
after installing the electronic components in the bottom layer,
and a second time once the sensors were installed.
In both cases isopropyl alcohol was used and the procedure took place in the clean room.
The boards were dried with nitrogen gas and attached to a metallic structure for proper handling.
The PCBs were tested and calibrated (see Section~\ref{sec:calib}),
and after a final cleaning procedure with alcohol and drying with nitrogen gas,
they were baked out at $80^{\, \circ}$C for 72 hours
and stored in antistatic bags ready for their transport and integration.

\section{Vacuum Compatibility Tests}
\label{sec:vacuum}

As presented above,
each half of the instrumented baffle accommodates a relatively large PCB with
various electronic components and 38 photodiodes,
as well as a wireless fidelity (WiFi) communication circuit with an antenna. 
Although most of the electronic components are compatible with UHV 
since they have ceramic and metallic encapsulations, 
some of the integrated circuits do not have their outgassing parameters
well defined or do not appear tabulated in the lists of NASA. 
To certify their use in a UHV environment, 
two independent studies were carried out, 
both resulting in very low outgassing and residual gas analyses with undetectable signal.

The vacuum compatibility of the different components of the baffle was first addressed at CERN 
by measuring their outgassing rate and possible transfer contamination. 
The outgassing rate was measured by the throughput method~\cite{outgassingRefe}. 
To perform outgassing rate measurements the samples were inserted into different cylindrical stainless-steel vacuum chambers 
whose dimensions depended on the shape of the samples; 
the smallest chamber was 60~mm in diameter and 200~mm high, 
while the largest one was 240~mm in diameter and 330~mm high. 
As the outgassing rate of the stainless steel chambers is at least one order of magnitude lower than that of the samples, 
except photo-sensors before bakeout, 
the different dimensions do not affect the results of the measurement. 
All vacuum chambers were equipped with a penning gauge and a residual gas analyser (RGA); 
they were pumped by a turbomolecular pump through an orifice providing a constant pumping speed;  
another penning gauge was installed near the pump aperture.

The different components of the baffle were split in four different batches: 
the gold-coated polyimide PCB, 
36 photo-sensors, 
the electronic components (resistors, condensers, microchips, etc.) 
and the connectors, antennae and wires. 
Along with the tested materials, 
witness samples were introduced in each batch to detect the presence of volatile condensable species released from the baffle elements. 
Such species, due to their high sticking probability, 
are adsorbed onto nearby surfaces escaping from the RGA detection. 
Typical volatile condensable species could contain, 
for example, 
solvents used during the polymer precursors production or the curing process, 
cleaning agents adsorbed onto the polymer matrix, 
and contaminants deposited on surfaces non accessible to cleaning. 
To identify the volatile condensable contamination on the witness samples, 
Fourier-transform infrared spectroscopy (FTIR) was applied following the procedure proposed by LIGO~\cite{outgassingRefe} and described in Ref.~\cite{LIGO_FTIR}. 
The objective of the FTIR measurements is to find the presence of condensable contaminants and possibly identify their nature.

The outgassing rate and residual gas composition were measured both at room temperature after 24 h of pump down 
and after a bakeout at $100^{\circ}$C during 24 h. 
The PCB was also tested after a bakeout at $100^{\circ}$C during 72 h. 
The measured outgassing rates are summarised in table~\ref{tab:outgassingCERN}.
Finally,
residual gas analysis did not show any contaminants above background levels during all measurements,
and FTIR analysis did not present any transfer of contaminants on the witness samples~\cite{FTIR_CERN}.

\begin{table}[htb]
\caption{\small
$Q_{24}$ outgassing rates as nitrogen equivalent in mbar~l/s of the different samples at room temperature before (24 h pumping) and after bakeout.}
\begin{center}
\begin{small}
\begin{tabular}{l c c c } 

Components        &  Unbaked     &  24 h bakeout    &    72 h bakeout    \\
\hline 
PCB               &  $1.3\times 10^{-5}$  & $8.2\times 10^{-7}$  &   $3.8\times 10^{-7}$ \\
Photodiodes       &  $9.4\times 10^{-8}$  & $5.6\times 10^{-9}$  &    \\
Components        &  $1.5\times 10^{-6}$  & $2.1\times 10^{-9}$  &    \\
Connectors, etc   &  $2.8\times 10^{-5}$  & $3.6\times 10^{-7}$  &    \\
\hline

\end{tabular}
\end{small}
\end{center}
\label{tab:outgassingCERN}
\end{table}

The assembled baffle was then tested at EGO to ensure compliance with Virgo UHV requirements.
In particular,
the points of concern were: 
\begin{itemize}
\item The soldering process used to integrate the electronic components to the PCB, 
due to the possible presence of high vapor pressure alloys not compatible with high temperature operations,
like the bake-out,
and also due to the possible presence of low-volatile additives (anti-oxidant fluxes), 
which are difficult to remove.
\item Some of the circuit components,
which are encapsulated in small plastic-like packages, 
when a ceramic alternative was not possible. 
\item The PCB itself, 
made of superposed layers of polyimide and copper, 
as a potential trap for contaminants and gases, mostly water vapor.
\end{itemize}

The UHV requirements to be met in the mirror cavities are: 
\begin{itemize}
\item Residual gas total flow below $10^{-4}$~mbar l/s. 
This requirement relates to the limit on the noise induced by the residual gas on the interferometer.
\item Partial pressure level of low-volatile molecules 
(so-called hydrocarbons) 
below $10^{-13}$~mbar.
This requirement is designed to avoid contamination of the optical surfaces due to condensation of hydrocarbons present in the residual gas. 
\end{itemize}

The total gas load measured with the RGA resulted in few $10^{-6}$~mbar~l/s after two weeks of pumping.
The outgassing rate of water vapor was relatively large,  
estimated to increase the total pressure in the vacuum cavity by few $10^{-9}$~mbar,
showing the usual decreasing trend controlled by bulk or surface effects becoming dominant at different stages of the process. 
The hydrocarbons’ partial pressure level was found to be of the order of $10^{-13}$~mbar after a mild and prolonged baking cycle. 

The absence of potential pollutants on the surfaces of the baffle was tested by FTIR and scanning electron microscopy  
with energy dispersive spectroscopy.
The inspection resulted in
no significant organic residuals and
the presence of few expected  metallic elements 
(Au+Ni, Cu, In+Sn, and Pb) and phosphorus
in acceptable rates.
Finally, 
the baffle was exposed to sample optics in-vacuum to verify the absence of optical losses related to contamination effects not detectable by RGA.
Tests performed at Laboratoire des Mat\'eriaux Avanc\'es (LMA)~\cite{LMA} did not show any significant increase in the contamination of the optics and 
the pollution remained at low level.

\section{Data Acquisition}
\label{sec:daq}

The DAQ system is divided into several layers of software,
which are implemented in different physical devices. 
On the lower level and sitting on one of the PCBs,
there is a microprocessor,
which we call the controller,
that governs the data taking 
and implements the communication channels with the upper layers. 
The controller is powered by an Ethernet controllable power supply (TTiPL-068P)
by means of \textsc{pyttilan}~\cite{pyttilan},
a separate library which allows to set the desired output voltage and current limit values,
and to monitor the actual values of the output.
The controller communicates with a server machine located outside of the IMC end mirror cavity
via two cables along the super attenuator up to the feedthrough located at the top of the IMC end mirror cavity.
Two additional cables are used to send the reset and boot signals to reprogram the baffle.
These cables are several metres long, 
which limits the maximum baud rate of serial communication to 38,400 bps.

\begin{figure}[htb]
\begin{center}

 \includegraphics[width=0.4\textwidth]{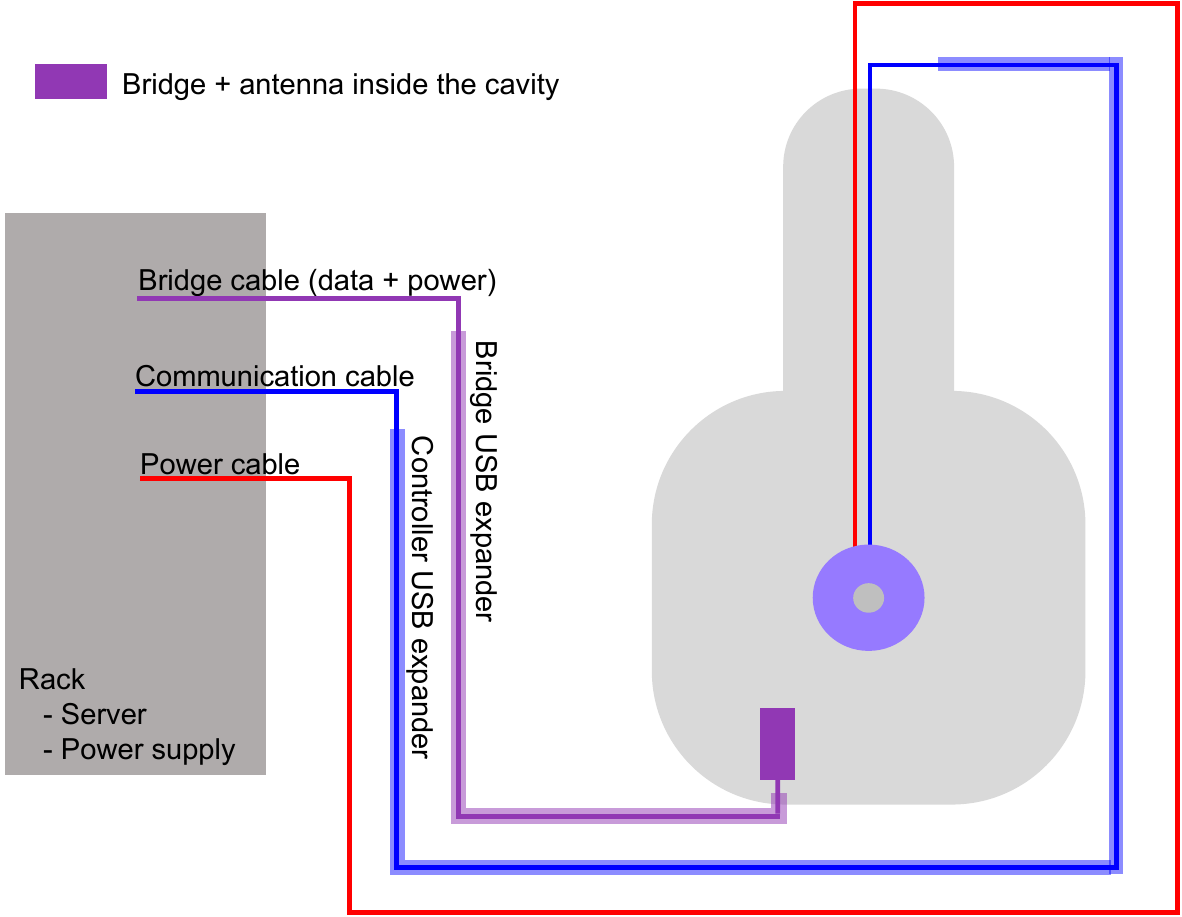}

\end{center}
\caption{\small
Scheme of the DAQ system of the instrumented baffle.
}
\label{fig:DAQscheme}
\end{figure}

On the other hand, 
since the baffle is placed inside the IMC end mirror cavity, 
which acts as a Faraday cage, 
a transmitter,
which we call the bridge,
is placed inside the IMC end mirror cavity with a direct view of the baffle,
to allow WiFi communication. 
The bridge is located near the feedthrough at the bottom of the IMC end mirror cavity,
and its only mission is to pass information between the controller and the server,
converting it from WiFi to serial and vice versa. 
Since the cables from the bridge to the lower feedthrough are much shorter than the ones from the controller to the top of the IMC end mirror cavity, 
one can reach much higher transmission rates with the former. 
Both the controller and bridge are based on an ESP32 chip,
which is a two-core microprocessor with 4~Mbytes of embedded flash memory externally extendable, 
and a seven-stage pipeline to support a clock frequency of up to 240~MHz.
Figure~\ref{fig:DAQscheme} schematises the different connections from the baffle to the outside server.

The server is in charge of sending commands to the controller and to receive its data. 
It is a Dell EMC PowerEdge R240 machine with an Intel Xeon E-2234 processor of 4.8 GHz, 
which contains the baffle DAQ server and implements the engineering database
and the main interfaces: 
the Virgo frame distribution and the graphical user interface (GUI).
The baffle DAQ server is the piece of software that operates and stores the baffle data, 
and is responsible for the coordination between the power supply and the controller,
as well as the different software subsystems.
Depending on the configuration, 
sent commands and data packages will go through the bridge or directly to the controller. 
The baffle DAQ server is implemented in \textsc{Python3} and uses \textsc{zeromq} as a protocol to communicate with its clients.

Both the ESP32 client and the power supply receive data from the devices underneath,
and the system is reactive to these data by a system of queues and threads.
A callbacks module executes different actions depending on the type of data received.
This architecture allows the implementation of several safety controls. 
In particular, 
the callbacks module checks the values of the ADC temperatures and if they are above a threshold value
the power supply turns off the controller.
Similarly,
if the communication is lost due to transmission errors, 
the change of state of the controller is detected and the communication is re-enabled, 
thus minimising the loss of data.

The eight ADCs of each board
are connected to the controller via one inter-integrated circuit (i2c) bus. 
The maximum frequency supported by the ADC-i2c interface in fast mode is 400~kHz,
which sets a limit to the serial clock.
This frequency gets highly reduced as
there are 38 channels enabled and 16 bits are required to send the value 
on the i2c bus.
The ESP32 chips are programmed using the \textsc{espressif} ESP-IDF~\cite{espressif},
which is based on \textsc{FreeRTOs}~\cite{freertos}. 
Since the ESP32 chips have two cores,
some of the tasks can be parallelised.  
The main tasks are: 
read the ADCs sensors, 
aggregate the read data in time windows, 
and send these aggregated data out of the system through a data structure called a frame.
Each frame includes the timestamp of the first sample, 
the timestamp of the last sample and 
the number of samples accumulated during that time window,
which are common to all channels.
For each channel enabled, 
the minimum,
the maximum,
and the sum of all the samples acquired during the time window are recorded.
On average, 
when the system is configured to provide frames at a rate of 2~Hz, 
each update contains the values of approximately 55~samples per channel.
In addition,
the temperatures of the ADCs are saved,
but since reading ADC temperatures is a slow operation,
it is done at a much lower frequency.

Finally,
at the top level of the instrumented baffle software are the operator interfaces. 
A command line interface was developed to allow access to the baffle server
via a secure shell (ssh) connection from any computer on the same network.
Its main features are: 
to view the state of the system; 
to establish the connection of the DAQ server to the controller or the bridge;
to control the power supply; 
to start or stop the data acquisition;
to visualise the last values received; 
and to save data to a disk.
In addition to the command line interface,
an engineering GUI is implemented in \textsc{Python} and \textsc{Qt}, 
which allows the use and validation of the system both during its development and commissioning phase. 
This GUI allows users to control the baffle DAQ and visualise the data received.
For example,
in the configuration screen the operator can select which ADC channels are enabled
and which are not. 
Being limited by the i2c bus bandwidth,
disabling some channels or full ADCs allows the number of samples acquired on each
report to increase, 
thus increasing the statistics for the enabled channels.
Although this is an expert mode interface, 
it includes some features that allow less trained people to use the baffle 
in a laboratory environment.
The engineering GUI and the client that integrates the instrumented baffle data in the Virgo DAQ 
use the \textsc{zeromq} interface to communicate with the baffle DAQ server.

\begin{figure}[htb]
\begin{center}

 \includegraphics[width=0.4\textwidth]{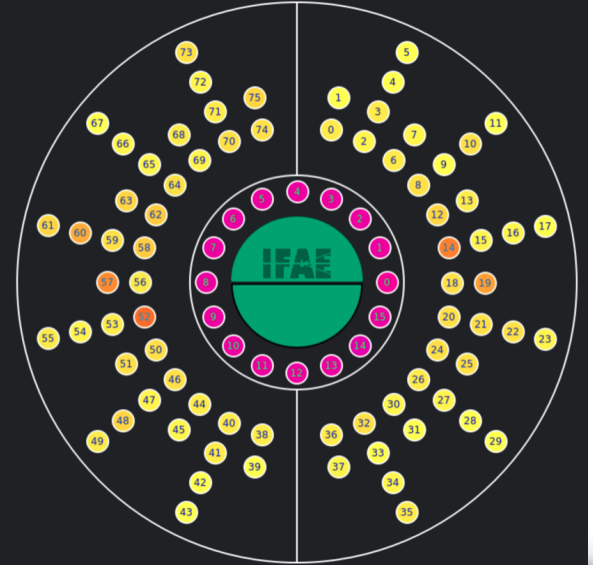}

\end{center}
\caption{\small
GUI for the operation of the IMC end mirror instrumented baffle.
The central green button turns the baffle on and off with a click.
The purple circles show the temperature of the 16 ADC,
and the readings from all 76 photodiodes are displayed in a color proportional to the ADC counts.
}
\label{fig:baffleGUI}
\end{figure}

A user friendly GUI was deployed in addition to the engineering GUI to facilitate 
the baffle operation.
With a simple click the baffle can be switched on and off and
the display shows the ADC counts together with the temperature readings,
which are updated with a frequency of 2~Hz.
This GUI,
shown in figure~\ref{fig:baffleGUI},
can be launched from any terminal with the baffle software installed 
and connected to the intranet of the EGO site.

\section{Photodiodes Calibration}
\label{sec:calib}

\begin{figure}[htb]
\begin{center}

 \includegraphics[width=0.4\textwidth]{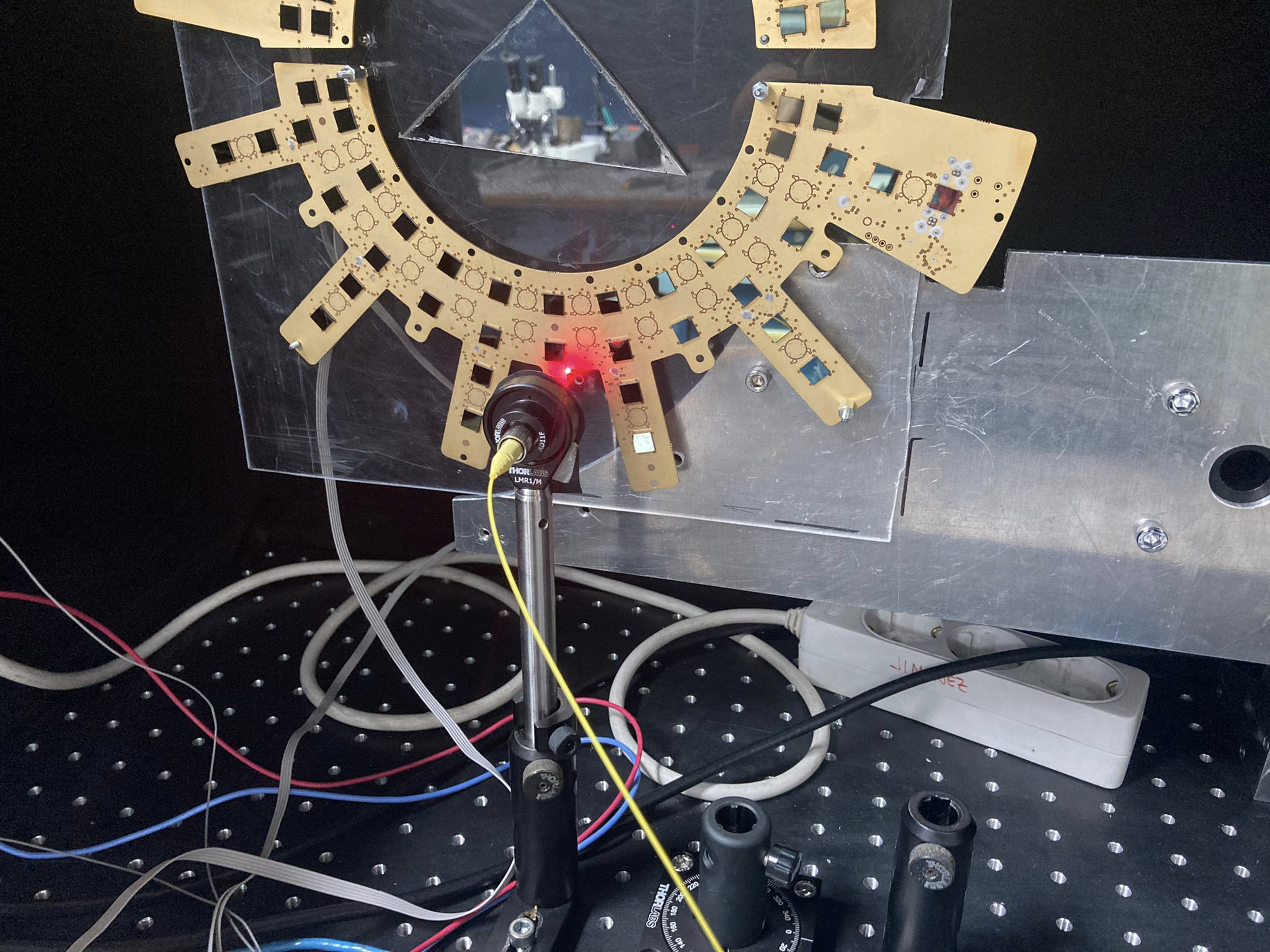}

\end{center}
\caption{\small
Optical setup used in the laboratory to calibrate the photodiodes.
A laser is attached to a collimator via an optical fibre.
This collimator directs the laser beam toward the photodiodes on the PCB.
The triangular mirror visible in the picture was used to help in the alignment of the laser.
}
\label{fig:calibrationSetup}
\end{figure}

A complete campaign of detector calibration took place before the installation of 
the instrumented baffle in the IMC cavity. 
Using a dedicated optical setup at Institut de F\'{\i}sica d'Altes Energies (IFAE),  
the detector response was characterized in terms of linearity, 
sensor-to-sensor variation and sensor absolute response. 
For the latter, 
a calibrated photo-sensor was used as reference,  
resulting on a value of 4.6~$\mu$W per ADC count,
with an uncertainty of about 5\%.
Figure~\ref{fig:calibrationSetup} shows the setup used in the laboratory to calibrate the sensors. 
In addition, the performance of the sensors was studied as a function of the temperature in the 
range between $23^{\rm o}$~C and $40^{\rm o}$~C observing no significant dependence.  

\begin{figure}[htb]
\begin{center}

 \includegraphics[width=0.47\textwidth]{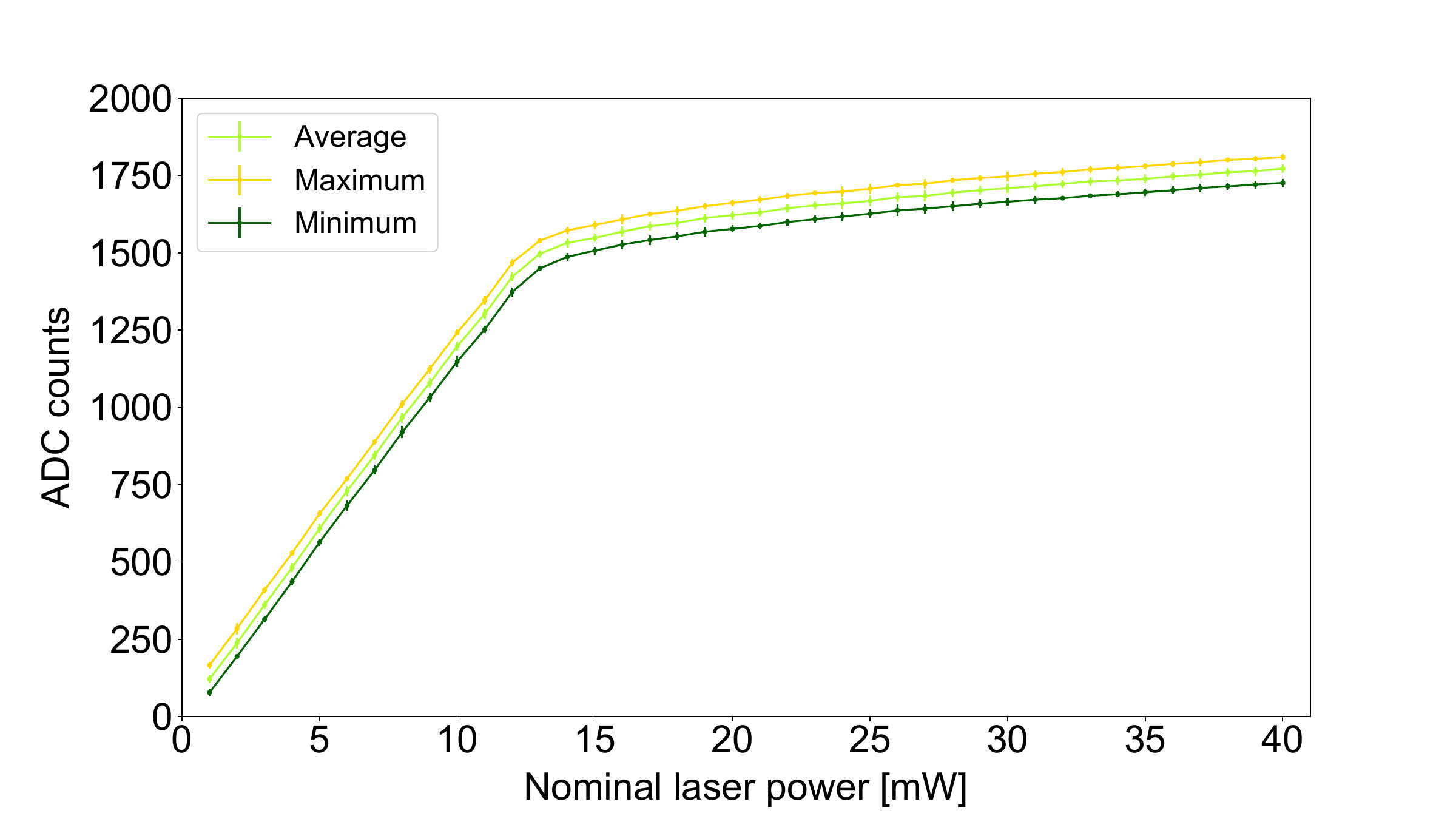}

\end{center}
\caption{\small
The minimum, the maximum and the average value of the ADC counts read in a period of 500~$\mu$s
for a given sensor are shown as a function of the nominal input power.
The measurements are repeated more than one hundred times for each laser power value.
}
\label{fig:sensorsResponse}
\end{figure}

In the optical setup, 
each of the PCBs with the sensors installed was mounted on an XY table facing the laser head inside a black box. 
Initially, 
a visible red laser was used to calibrate the XY table to ensure that the laser spot was well centered on each of the silicon sensors before operating the laser at 1064~nm. 
The laser signal itself was propagated inside the black box using optical fibres instrumented with collimators, 
limiting the laser transverse size to a diameter of 1.38~mm at a focal distance of about 6~mm. 
A fine XY scan across the surface of a single photo-sensor indicated that the response of the sensor does not depend on the exact location of the center of the laser provided the laser spot is fully contained in the detector active area.
The response of each of the individual sensors was studied as a function of the nominal laser power,
as shown in Figure~\ref{fig:sensorsResponse},
with nominal laser power values in the range between 1~mW and 40~mW.  

All the sensors presented a linear behaviour up to about 15~mW,  
when the sensor together with the DAQ electronics start showing saturation effects. 
The sensor-to-sensor variation in the response is  within a $3\%$ uncertainty,
which is propagated into the final absolute calibration of the sensors.   
The amount of light the sensors are exposed to is smaller than the nominal power displayed by the laser since significant losses take place in the optical fibres and collimators. 
In addition, 
the laser presents large fluctuations at very low laser powers.  
In order to determine the level of losses, 
a fully calibrated photo-sensor was used.  
The measurements indicate that the laser losses are $54.1 \pm 1.6 \%$, 
where the quoted uncertainty includes both statistical and experimental uncertainties.

\section{Baffle Installation}
\label{sec:install}

\begin{figure}[htb]
\begin{center}

 \includegraphics[width=0.4\textwidth]{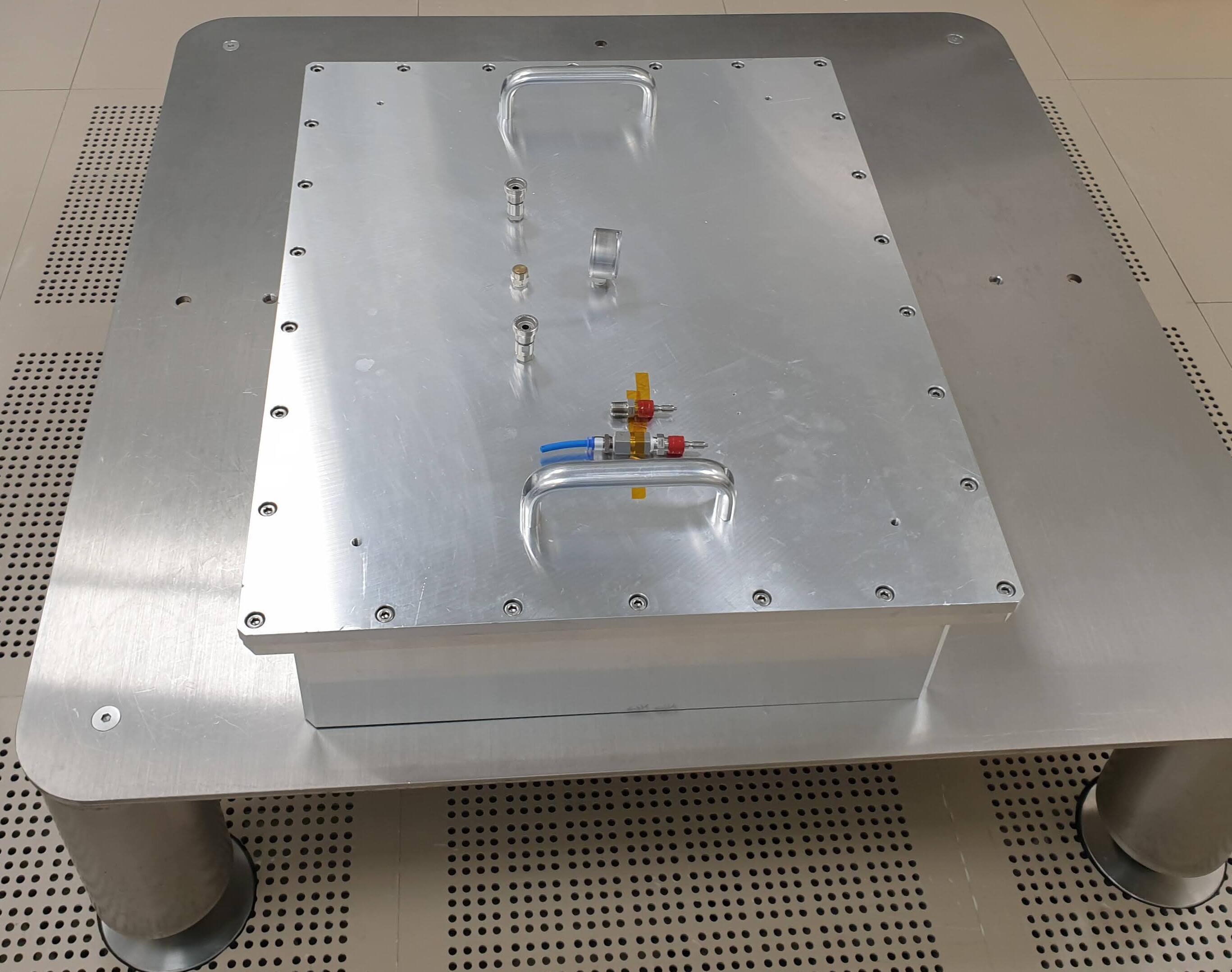} 
 \includegraphics[width=0.4\textwidth]{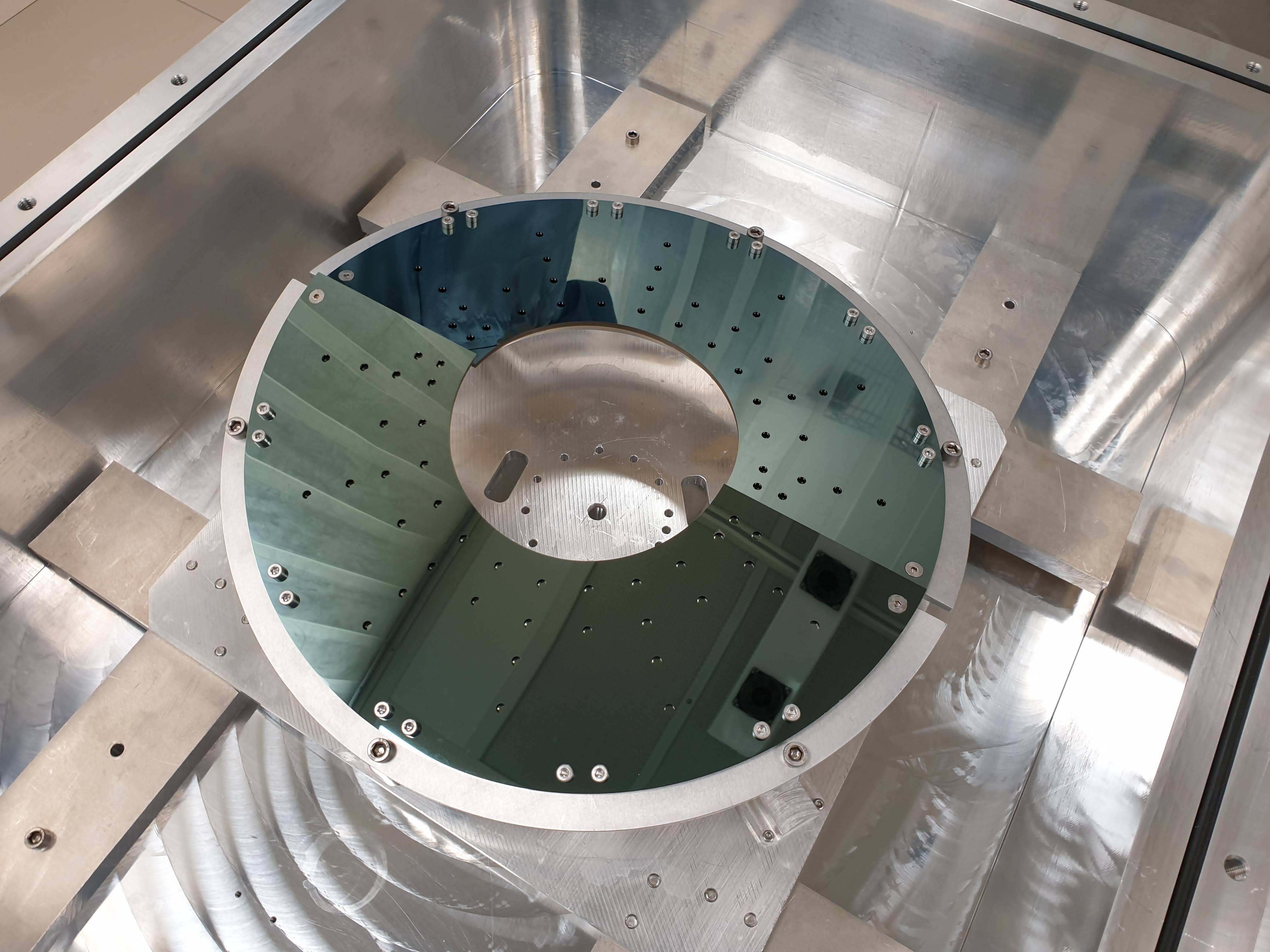} 

\end{center}
\caption{\small
(top) Transportation box containing the instrumented baffle.
(bottom) The instrumented baffle before its installation inside the transport box.
}
\label{fig:newBaffleBefore}
\end{figure}

The PCBs were attached to the baffle in the clean room at IFAE,
and the ready-to-install instrumented baffle was bolted to an aluminum transporting box.
This box was filled with nitrogen gas with an over-pressure of 0.6~bar to preserve the baffle in a non-reactive atmosphere.
The supports in the box incorporate springs to isolate the baffle from possible external
disturbances during its transportation to the EGO site.
Two spare PCBs,
and the support shown in Fig.~\ref{fig:baffleMold},
which had been cleaned in an ultrasonic bath with a solution of citranox at 10\% during ten minutes,
rinsed with deionised water and
dried during 48 hours at 100$^{\,\circ}$C,
were sealed inside antistatic bags
and also shipped to EGO.
Figure~\ref{fig:newBaffleBefore}~(top) shows the transportation box with the baffle inside
at its arrival to EGO,
and figure ~\ref{fig:newBaffleBefore}~(bottom) shows the baffle inside the box,
ready for installation.

Once at EGO,
the baffle was removed from the box and attached to the support.
This operation took place in the clean room.
After checking its correct functioning,
the two half-baffles were protected inside sealed bags and moved to the IMC end mirror cavity.

\begin{figure}[htb]
\begin{center}

\includegraphics[width=0.4\textwidth]{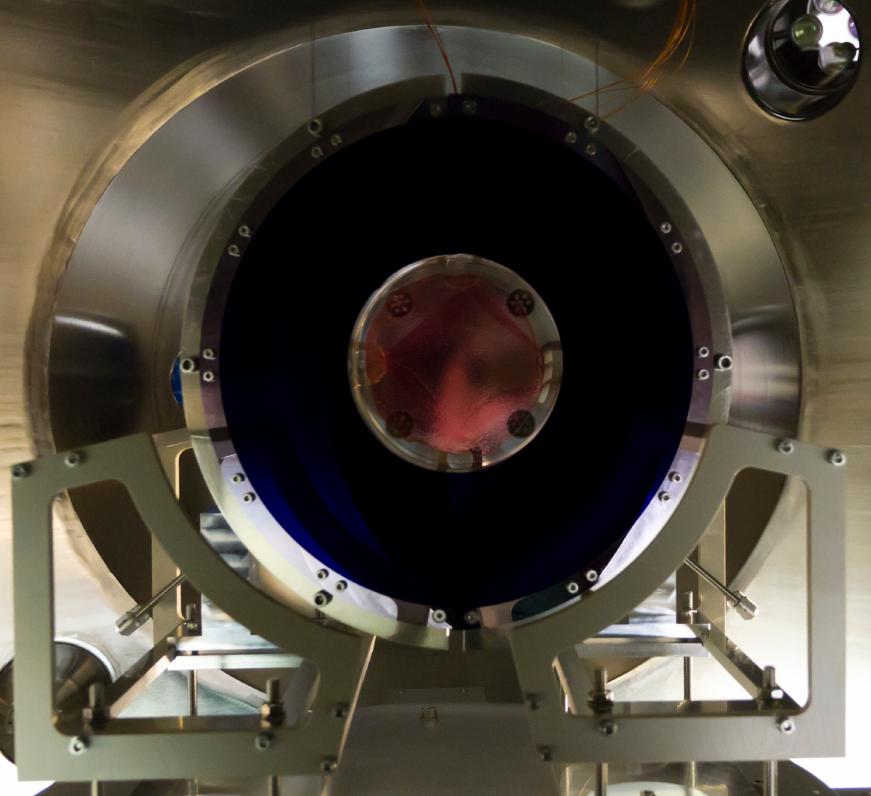}
\includegraphics[width=0.4\textwidth]{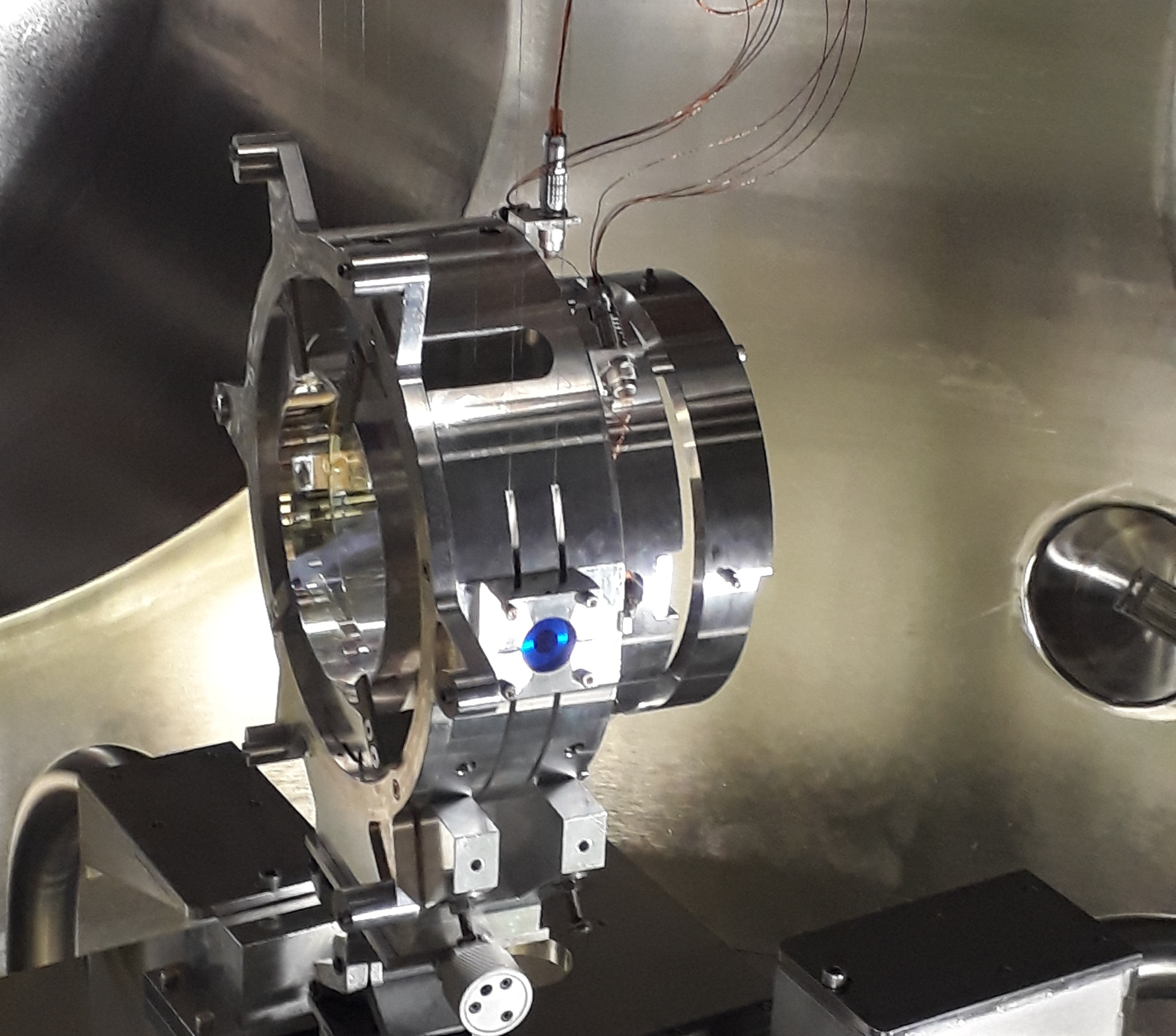}

\end{center}
\caption{\small
(top) Old non-instrumented baffle before being removed from the payload.
(bottom) Supporting structure ready for the installation of the instrumented baffle.
The installation took place in the IMC end mirror cavity.
}
\label{fig:baffleSupport}
\end{figure}

The replacement of the old baffle by the instrumented one was a delicate operation
that took place inside the IMC end mirror cavity in April~2021.
On Saturday, April~24,
the local controls of the IMC suspensions were opened and
the superattenuator control was left in the DC position
to allow the venting of the IMC end mirror cavity,
which proceeded during the weekend.
On Monday,
April~26,
the IMC end mirror cavity was opened and cleaned,
the marionette was fixed to the safety frame,
and the payload reference mass was clamped to the custom cradle.
Finally, 
the old baffle was dismounted.
Figure~\ref{fig:baffleSupport} shows the payload before and after
removing the old baffle.
The same supporting structure was used to attach the instrumented baffle.

\begin{figure}[htb]
\begin{center}

 \includegraphics[width=0.4\textwidth]{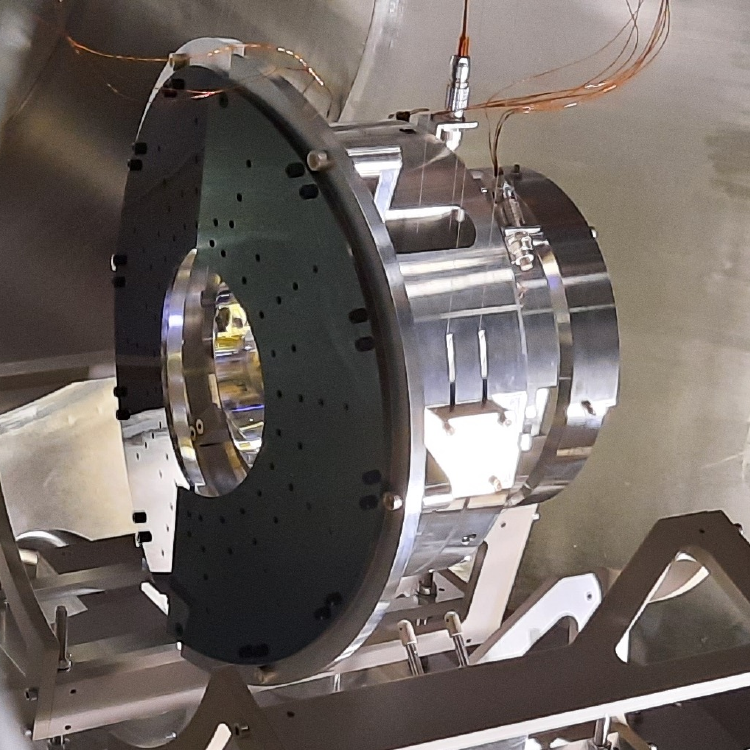}

\end{center}
\caption{\small
Instrumented baffle after its installation in the payload of the IMC end mirror.
}
\label{fig:NewBaffle}
\end{figure}

In parallel, 
the functionality of the instrumented baffle was tested first in
the clean room and then outside of the IMC end mirror cavity,
before moving it to its final place.
The server was installed in one of the racks close to the IMC end mirror cavity,
and the readout and power cables connected in both ends.
On Wednesday,
April~28, 
before closing the IMC end mirror cavity,
the correct operation of the baffle was verified.
In particular,
the sensors were illuminated one by one with a green laser to confirm that
they were all operational.
Although the WiFi communication system was tested to work correctly outside the IMC end mirror cavity,  
as well as inside a separate vacuum vessel at EGO, 
it was not possible to fully commission it once the baffle was installed inside the IMC end mirror cavity. 
Further investigations to recover the WiFi readout would have required interventions in the boot line affecting pre-installed cables inside the IMC end mirror cavity. 
Such interventions were considered too risky to preserve the integrity of the IMC payload. Figure~\ref{fig:NewBaffle} shows the instrumented baffle once installed in front of the
IMC end mirror.

\section{Instrumented Baffle Operation}
\label{sec:performance}

As mentioned above, 
the instrumented baffle was installed in front of the end mirror of the IMC cavity in April 2021 
and has been collecting data continuously since its installation.
The saved data include not only the ADC counts of the 76 photodiodes, 
but also the values of the 16 temperature sensors and
the voltage and current that the power supply provides to the electronics.
These data are stored as CSV files in a local server, 
to allow quick performance studies,
as well as in an \textsc{InfluxDB} database~\cite{influxdata},
for easier access and real-time data visualisation with \textsc{grafana}~\cite{grafana}. 

\begin{figure}[htb]
\begin{center}

   \includegraphics[width=0.47\textwidth]{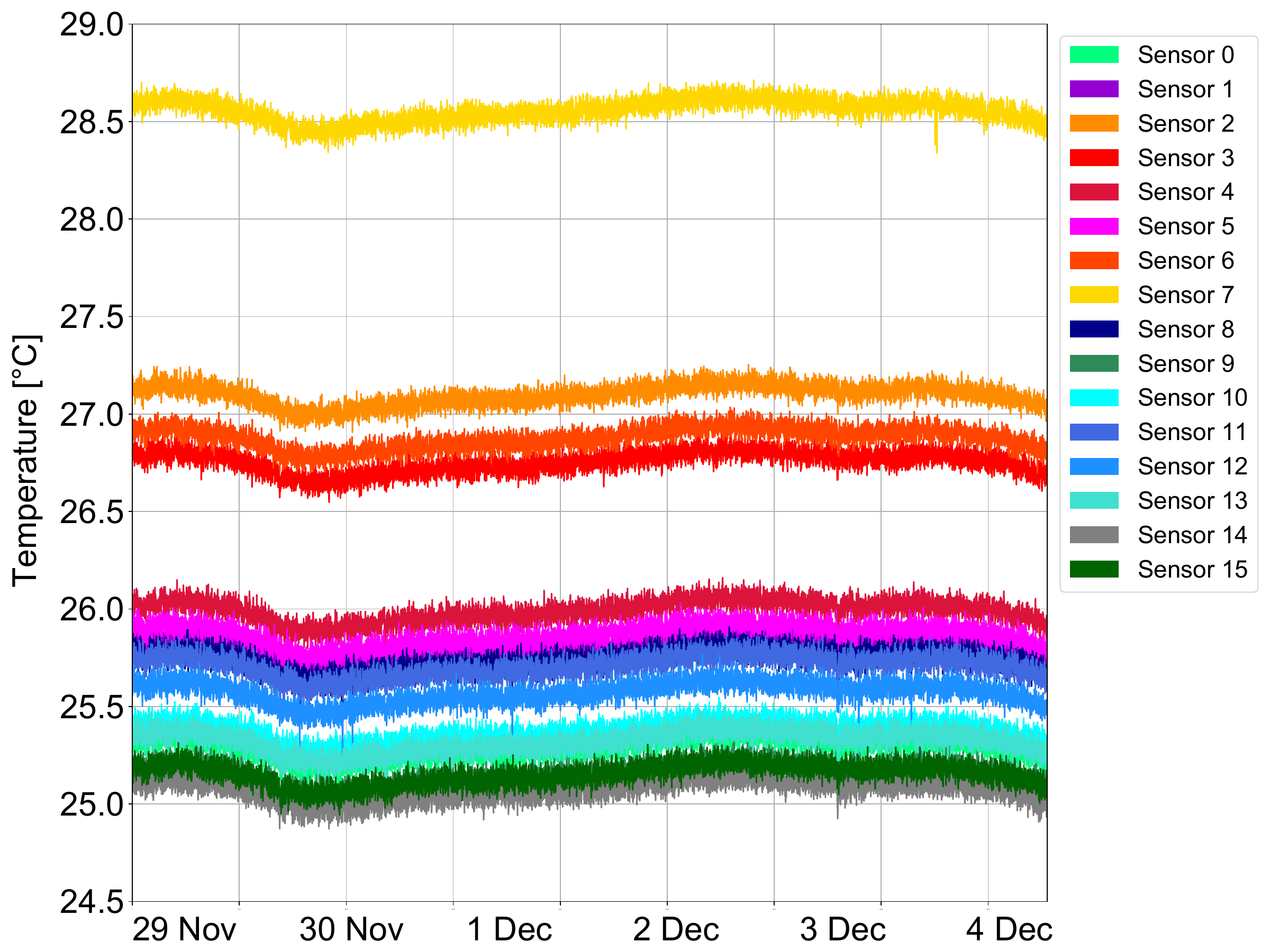} 

\end{center}
\caption{\small
Temperature evolution of the 16 sensors for a period of five days.
The sensor with the highest readings corresponds to that closest to the controller,
dissipating most of the heat.
}
\label{fig:temperatureWekly}
\end{figure}

\begin{figure}[htb]
\begin{center}

   \includegraphics[width=0.47\textwidth]{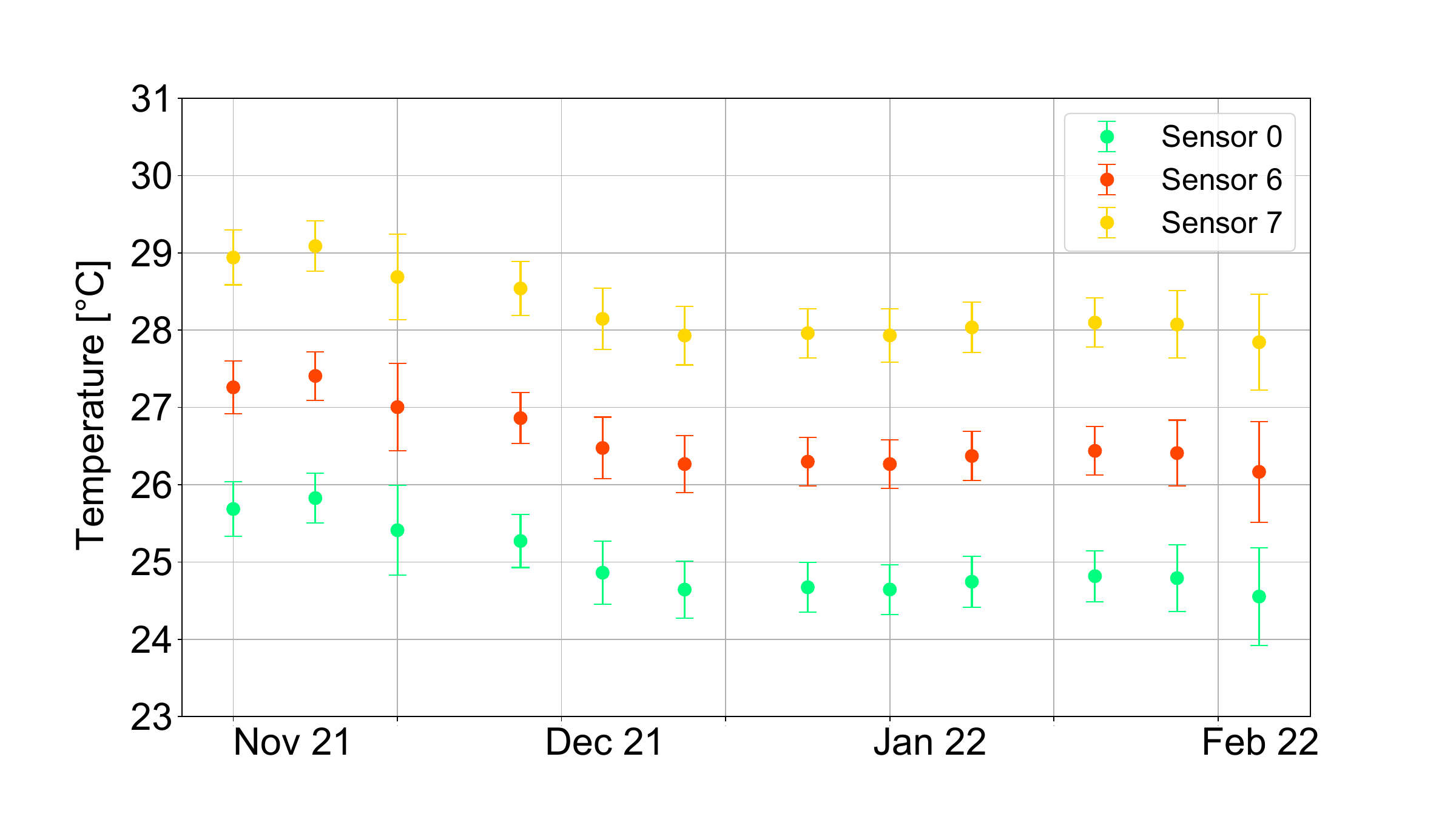} 

\end{center}
\caption{\small
Temperature evolution of three representative temperature sensors for a period of four months.
The sensor with the highest readings corresponds to that closest to the controller,
dissipating most of the heat.
}
\label{fig:temperatureMontly}
\end{figure}

\begin{figure}[htb]
\begin{center}

   \includegraphics[width=0.47\textwidth]{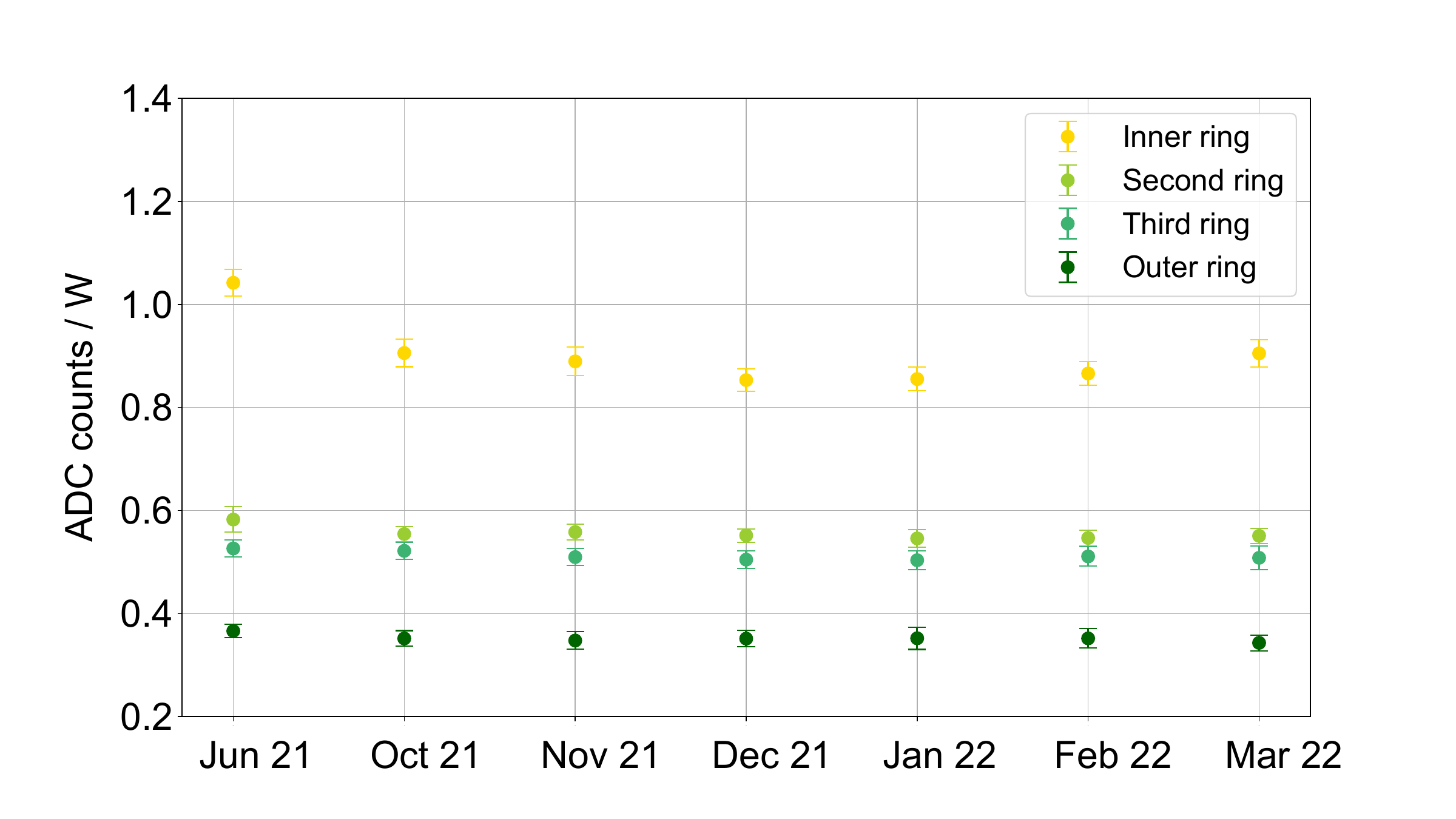} 

\end{center}
\caption{\small
Pedestal subtracted sensor output (normalised to the injection laser power) averaged for
each of the four rings of sensors for a period of nine months.
The sensors are grouped into rings according to their distance to the center of the mirror.
}
\label{fig:stability}
\end{figure}

The suspended baffle operates at room temperature and under HV conditions, 
without any cooling system due to the limitations dictated by the suspension.
So far no overheating has been observed thanks to 
efficient heat dissipation from the gold-plated PCB,
a careful design of the mechanical couplings with the stainless steel structure 
and a moderate operating tension.
In any case, 
there is a protection system that shuts down the baffle if any of the ADC 
temperatures exceeds $35^{\,\circ}$~C.
Figures~\ref{fig:temperatureWekly}~and~\ref{fig:temperatureMontly} show the weekly and monthly
evolution of the temperature, 
respectively.
The sensor showing the highest temperature is the one closest to the controller. 
Since its commissioning,
the data taken by the baffle show good stability,
as can be seen in figure~\ref{fig:stability}.

\begin{figure}[htb]
\begin{center}

   \includegraphics[width=0.47\textwidth]{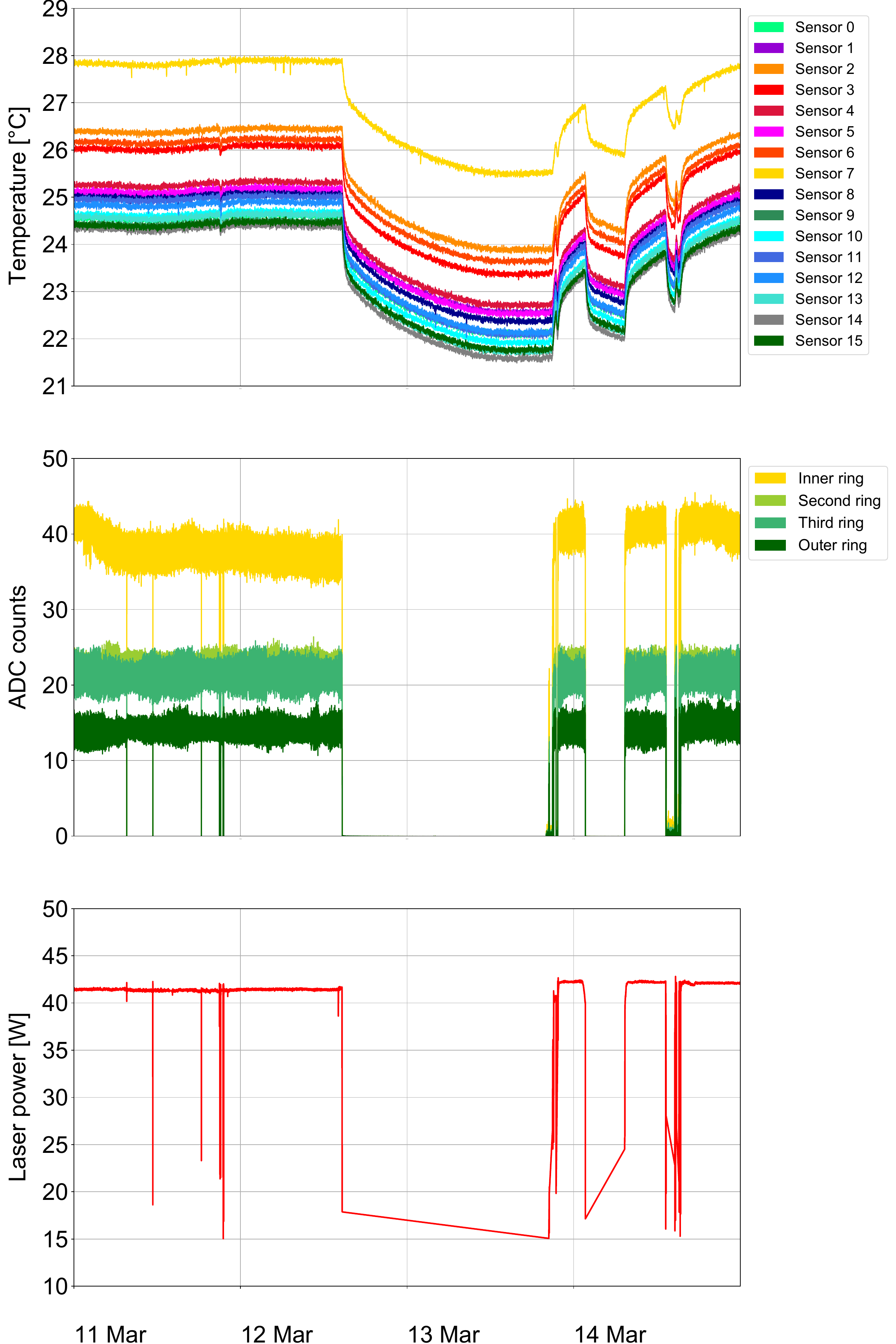} 

\end{center}
\caption{\small
(top) Baffle temperatures,
(middle) average sensor signals for each of the four rings,
and (bottom) input laser power over a four-day period.
}
\label{fig:VirgoData}
\end{figure}

Finally, 
both the sensor and the temperature
read-out values are written in the standard Virgo data-stream, 
allowing for more general studies, 
including correlation with other Virgo variables such as laser power or seismic status.
Figure~\ref{fig:VirgoData} presents the laser power,
the sensor signals averaged for each of the four rings,
and the baffle temperatures over a four-day period,
showing a clear correlation between laser power and baffle readings.

\section{Summary and Outlook}
\label{sec:sum}

In this paper, we have presented a novel technology
aimed to instrument with photo-sensors the baffles that surround the 
suspended mirrors acting as test masses in the Virgo interferometer. This new device has the potential to improve the understanding of the stray light distribution at low scattering angles inside the optical cavities, to help detect the appearance of higher order modes, to 
monitor the contamination of the mirror surfaces, and to  facilitate the pre-alignment of the interferometer.

A prototype  was built and installed in front of the end mirror of the 
Virgo interferometer input mode cleaner in the spring of 2021.
The baffle has been taking data smoothly since the installation 
and the data taken,
ADC counts and temperatures,
show good stability and a clear correlation between laser power and baffle readings is observed.
This baffle,
which will be operating in the upcoming O4 observation run,
serves as a demonstrator of the technology  being now used to construct 
new instrumented baffles, with larger number of sensors and improved optical properties, 
for Virgo’s main mirrors in time for the O5 observation run.

\section*{Acknowledgements}

The  authors  gratefully  acknowledge  the  European  Gravitational Observatory (EGO) and the Virgo Collaboration for providing access to the facilities 
and the support from EGO and the University of Pisa during the installation of the new instrument in Virgo; 
CERN for their contribution to the vacuum studies;
the Optical Institute OI-CSIC in Madrid for their assistance in the reflectance measurements;
Sensofar Metrology for providing roughness measurements of the baffle surface;
and ALBA Synchrotron for their technical support.
This work is partially supported by the Spanish MCIN/AEI/10.13039/501100011033 under
the grants SEV-2016-0588, PGC2018-101858-B-I00,  
and PID2020-113701GB-I00,
some of which include ERDF  funds  from  the  European  Union. 
IFAE  is  partially funded by the CERCA program of the Generalitat de Catalunya.


\bibliographystyle{apsrev}
\bibliography{baffle}{}

\end{document}